\let\latexarabic\arabic
\let\latexdocument\document
\let\latexenddocument\enddocument

\RequirePackage[thmmarks]{ntheorem}
\makeatletter
\renewtheoremstyle{plain} 
  {\item[\hskip\labelsep \theorem@headerfont ##1\ \textup{##2}\theorem@separator]} 
  {\item[\hskip\labelsep \theorem@headerfont ##1\ \textup{##2}\ (##3)\theorem@separator]}
\makeatother

\documentclass[
  lineno
]{biometrika}

\let\document\latexdocument
\let\enddocument\latexenddocument
\AtEndDocument{\printhistory}
\let\arabic\latexarabic
\def\rm{}

\usepackage{amsmath}

\usepackage{times}
\usepackage{bm}
\usepackage{natbib}

\usepackage[plain,noend]{algorithm2e}

\let\originalleft\left
\let\originalright\right
\renewcommand{\left}{\mathopen{}\mathclose\bgroup\originalleft}
\renewcommand{\right}{\aftergroup\egroup\originalright}

\makeatletter 
\renewcommand{\algocf@captiontext}[2]{#1\algocf@typo. \AlCapFnt{}#2} 
\def\@algocf@capt@plain{top}
\renewcommand{\algocf@makecaption}[2]{%
  \addtolength{\hsize}{\algomargin}%
  \sbox\@tempboxa{\algocf@captiontext{#1}{#2}}%
  \ifdim\wd\@tempboxa >\hsize
    \hskip .5\algomargin%
    \parbox[t]{\hsize}{\algocf@captiontext{#1}{#2}}
  \else%
    \global\@minipagefalse%
    \hbox to\hsize{\box\@tempboxa}
  \fi%
  \addtolength{\hsize}{-\algomargin}%
}
\makeatother

\usepackage{xr}
\usepackage{amssymb} 
\usepackage{url} 
\usepackage{makecell} 

\def\T{{ \mathrm{\scriptscriptstyle T} }} 

\newcommand{\ee}{E}

\newcommand{\cov}{\mathrm{cov}}
\newcommand{\var}{\mathrm{var}}

\newcommand{\pr}{\mathrm{pr}}

\DeclareMathOperator{\gtr}{tr}

\newcommand{\N}{N}
\newcommand{\Ga}{\textnormal{Ga}}

\newcommand{\law}{\Pi}

\newcommand{\Sec}{$\S\,$}

\begin{document}



\markboth{W. van den Boom, G. Reeves \and D. B. Dunson}{Integrated rotated Gaussian approximation}

\title{Approximating posteriors with high-dimensional nuisance parameters via integrated rotated Gaussian approximation}

\author{W. VAN DEN BOOM}
\affil{Yale-NUS College, National University of Singapore, 16 College Avenue West \#01-220, Singapore 138527, Singapore\email{willem@yale-nus.edu.sg}}

\author{G. REEVES \and D. B. DUNSON}
\affil{Department of Statistical Science, Duke University, Box 90251, Durham,\\ North Carolina 27708, U.S.A. \email{galen.reeves@duke.edu} \email{dunson@duke.edu}}

\maketitle

\begin{abstract}
Posterior computation for high-dimensional data with many parameters can be challenging.
This article focuses on a new method for approximating posterior distributions of a low- to moderate-dimensional parameter in the presence of a high-dimensional or otherwise computationally challenging nuisance parameter.
The focus is on regression models and the key idea is to separate the likelihood into two components through a rotation.
One component involves only the nuisance parameters,
which can then be integrated out using a novel type of Gaussian approximation.
We provide theory on approximation accuracy that holds for a broad class of forms of the nuisance component and priors.
Applying our method to simulated and real data sets shows that it can outperform state-of-the-art posterior approximation approaches.
\end{abstract}


\section{Introduction}

Consider the regression model
\begin{equation} \label{eq:lm}
	y \sim \N\left(X \beta + \eta,\ \sigma^2 I_n \right),
\end{equation}
where $y$ is an $n$-dimensional vector of observations, $X$ is an $n\times p$ design matrix, $\beta$ is a $p$-dimensional parameter of interest, $\eta$ is an $n$-dimensional nuisance parameter, and $\sigma^2$ is the error variance.
The nuisance parameter can for instance capture the effect of a large set of covariates not included in $X$,
or of non-Gaussian errors.
Our goal is Bayesian inference on the model in \eqref{eq:lm} when $p$ is of moderate size such that $p\ll n-p$ with the focus on the posterior
\begin{equation} \label{eq:postB}
	\pi(\beta\mid y)
	= \int \pi(\beta,\eta \mid y)\,d\eta
	= \frac{1}{\pi(y)} \int \pi(y\mid\beta,\eta)\, \pi(\beta,\eta) \,d\eta.
\end{equation}
The integrals in \eqref{eq:postB} and $\pi(y)$ are intractable to approximate accurately for certain priors $\pi(\beta,\eta)$, with direct approximations such as Laplace's method producing inaccurate results and Monte Carlo sampling being daunting computationally.
Our key idea is to transform the hard problem with nuisance parameter $\eta$ in a principled way to a $p$-dimensional one which can be written as a linear model including only $\beta$. Then, a low-dimensional inference technique can be applied to this $p$-dimensional model.
The transformation uses a novel type of Gaussian approximation using a data rotation to integrate out $\eta$ from \eqref{eq:lm}.

Section~\ref{sec:examples} discusses special cases of the model in \eqref{eq:lm}.
Applications include epidemiology studies in which $y$ is a health outcome, $X$ consists of exposures and key clinical or demographic factors of interest, and $\eta$ is the effect of high-dimensional biomarkers.
The goal is inference on the effect of the exposures and the clinical or demographic covariates, but adjusting for the high-dimensional biomarkers.
For example, $\eta$ may result from genetic factors, such as single-nucleotide polymorphisms (SNPs), and we want to control for these in identifying an environmental main effect. It is often impossible to isolate the impact of individual genetic factors
so we consider these effects as nuisance parameters.
Another use of \eqref{eq:lm} is computation of posterior inclusion probabilities in high-dimensional Bayesian variable selection as detailed in \Sec{}\ref{sec:bvs}.

Data with a complex component $\eta$ that is not of primary interest and
only a moderate number $p$ of parameters of interest, are more and more common. Unfortunately, the complexity of $\eta$ can make accurate approximation of $\pi(\beta\mid y)$ in \eqref{eq:postB} challenging even when $p=1$.
One naive approach is to ignore the nuisance parameter $\eta$ by setting it to zero.
The result can be problematic as omitting $\eta$ changes the interpretation of the parameter of interest $\beta$, which therefore might take on a different value.
For example, $\eta$ might capture the effect of covariates with it being important to adjust for them to avoid misleading conclusions on $\beta$.

Many posterior approximation methods exist, including Monte Carlo \citep{George1993,George1997,Ohara2009}, variational Bayes \citep{Carbonetto2012,Ormerod2017}, integrated nested Laplace approximations \citep{Rue2009}, and expectation propagation \citep{HernandezLobato2014}. However, these methods can be computationally expensive, do not apply to our setting, or lack theoretical results regarding approximation accuracy.
A notable exception to the latter is the fast posterior approximation algorithm of \citet{Huggins2017} which comes with bounds on the approximation error under conditions on the prior such as log-concavity, Gaussianity, and smoothness.
The class of priors that we allow on $\beta$ and $\eta$ is much larger.
Our method and its analysis for instance apply to dimensionality reduction and shrinkage priors such as spike-and-slab, horseshoe, and Laplace distributions.

The main computational bottleneck of our method is calculation of the mean and variance of a nuisance term, for which one can choose any suitable algorithm. As a result, the computational cost of our method is comparable to that of the fast algorithm chosen for this step.

\section{Integrated rotated Gaussian approximation}
\label{sec:irga}

\subsection{Notation and assumptions}
\label{sec:setup}

Denote the multivariate Gaussian distribution with mean $\mu$ and covariance $\Sigma$ by $\N\left(\mu,\ \Sigma\right)$, and its density function evaluated at $a$ by $\N\left(a \mid \mu,\ \Sigma\right)$.
Denote the distribution of $a$ conditional on $b$ by $\law(a\mid b)$ and its density, with respect to some dominating measure, evaluated at $a$ by $\pi(a\mid b)$.
We assume that $\beta$ and $\eta$ are a priori independent so that $\law(\beta,\eta) = \law(\beta)\law(\eta)$.
We treat $X$ and $\sigma^2$ as known constants unless otherwise noted.
Assume that $p\leq n$.
We assume that $X$ is full rank to simplify the exposition, but our method also applies to rank deficient $X$.

\subsection{Description of the method}
\label{sec:dataRot}

We integrate out $\eta$ from \eqref{eq:lm} by splitting the model into two parts, one of which does not involve $\beta$. A data rotation provides such a model split. Specifically, consider as rotation matrix the $n\times n$ orthogonal matrix $Q$ from the QR decomposition of $X$. 
Define the $n\times p$ matrix $M$ and the $n\times(n-p)$ matrix $S$ by $\left({M},\ {S}\right) = Q$.
Then, the columns of ${M}$ form an orthonormal basis for the column space of $X$ since $X$ is full rank by assumption \citep[\Sec{}5.2]{Golub1996}.
Since $Q$ is orthogonal,
the columns of ${S}$ form an orthonormal basis for the orthogonal complement of the column space of $X$. Therefore, ${S}^\T X = 0_{(n-p)\times p}$, an $(n-p)\times p$ matrix of zeros, which can also be derived from the fact that $Q^\T X$ is upper triangular.

By the rotational invariance of the Gaussian distribution and $Q^\T Q = I_n$, ${Q^\T y} \sim {\N\left(Q^\T X\beta + Q^\T \eta,\ \sigma^2 I_n\right)}$ is distributionally equivalent to \eqref{eq:lm}. This rotated model splits as
\begin{subequations} \label{eq:rot}
\begin{align}
	\label{eq:rot1}
	{M}^\T y &\sim \N\left(
		{M}^\T X\beta
		+ {M}^\T \eta,\
		\sigma^2 I_p
	\right), \\
	\label{eq:rot2}
	{S}^\T y &\sim \N\left(
		{S}^\T \eta,\
		\sigma^2 I_{n-p}
	\right);
\end{align}
\end{subequations}
using ${S}^\T X = 0_{(n-p)\times p}$.
This transformation motivates a two-stage approach in which one first computes  $\law(\eta  \mid {S}^\T y  )$ from submodel \eqref{eq:rot2} and then uses this  distribution as an updated prior for the projected nuisance term $M^\T \eta$ in submodel \eqref{eq:rot1}. Following this approach,  the posterior of $\beta$ can be expressed as
$\law(\beta \mid y) \propto \Pi(\beta) \int \N( M^\T y \mid M^\T X \beta + M^\T \eta,\, \sigma^2 I_p) \, d \law( M^\T \eta \mid S^\T y)$.

In practice, $\law\left( {M}^\T \eta \mid {S}^\T y \right)$ may be intractable to compute exactly because of the complexity of $\law(\eta)$. To alleviate this challenge, we  consider an approximation $\hat{\law}\left( {M}^\T \eta \mid {S}^\T y \right)$, which then leads to an approximation for the posterior of $\beta$: 
\begin{align}
\hat{\law}(\beta \mid y) \propto \Pi(\beta) \int \N( M^\T y \mid M^\T X\beta + M^\T \eta,\, \sigma^2 I_p) \, d \hat{\law}(M^\T \eta \mid S^\T y) . \label{eq:ira}
\end{align}
All distributions, densities, and probabilities
resulting from this approximation carry a hat to distinguish them from their exact counterparts.

A Gaussian approximation is analytically convenient:
\begin{equation} \label{eq:gauss}
	\hat{\law}\left( {M}^\T \eta \mid {S}^\T y \right)
	= \N(\hat{\mu},\, \hat{\Sigma}),
\end{equation}
where $\hat{\mu}$ and $\hat{\Sigma}$ are estimates of the mean and covariance of $\law\left( {M}^\T \eta \mid {S}^\T y\right)$, respectively. In this case, \eqref{eq:ira} simplifies as
\begin{align}
\hat{\law}(\beta \mid y) \propto \Pi(\beta)\, \N( M^\T y \mid M^\T X \beta + \hat{\mu} ,\,   \sigma^2 I_p + \hat{\Sigma} ). \label{eq:irga}
\end{align}
Only $\beta$ is unknown such that the computational problems with \eqref{eq:lm} resulting from the complexity of $\law(\eta)$
have been resolved in \eqref{eq:irga}.
Furthermore, \eqref{eq:irga} is equivalent to a Gaussian linear model with observations ${M}^\T y - \hat{\mu}$,
design matrix ${M}^\T X$,
and parameter $\beta$.
We have reduced a model with a potentially challenging nuisance parameter to a low-dimensional one with the nuisance integrated out while controlling for the effect of the nuisance parameter in a principled manner.
Algorithm~\ref{alg:irga} summarizes our method when the Gaussian approximation from \eqref{eq:gauss} is used.

\begin{algo} \label{alg:irga}
Integrated rotated Gaussian approximation. \\
Input: Data $(y,\ X)$
\begin{enumerate}[1.]
		\item
		Compute the QR decomposition of $X$ to obtain the rotation matrix $Q = \left({M},\ {S}\right)$.
		\item
		Compute the estimates $\hat{\mu}$ and $\hat{\Sigma}$ for the mean and covariance of $\law\left( {M}^\T \eta \mid {S}^\T y\right)$ based on submodel \eqref{eq:rot2} using an algorithm of choice.
		\item
		Approximate the posterior $\law(\beta\mid y)$ according to \eqref{eq:irga}.
	\end{enumerate}
Output: The approximate posterior $\hat{\law}(\beta\mid y)$
\end{algo}

\subsection{Relation to other methods}

Algorithm~\ref{alg:irga} has resemblances with other approximation methods.
Integrated nested Laplace approximations \citep{Rue2009} also approximate a nested part of a Bayesian model by a Gaussian distribution
but with important differences.
A Laplace approximation is applied without a data rotation and is done at two, rather than one, nested levels of the model. Moreover, a Laplace approximation matches the mode and curvature of the approximating Gaussian while \eqref{eq:gauss} matches the moments.
Laplace's method \citep{Kadane1986} requires a continuous target distribution
and
integrated nested Laplace approximations assume a conditionally Gaussian prior on some parameters.
Our Gaussian approximation needs no such conditions on priors but assumes a Gaussian error distribution.
For instance, \Sec{}\ref{sec:examples} considers examples of priors on $\eta$ that are not continuous or are non-Gaussian.

The approximation in \eqref{eq:gauss} aims to match the first two moments of the exact $\law\left( {M}^\T \eta \mid {S}^\T y \right)$.
Such matching is the principle behind expectation consistent inference \citep{Opper2005}.
Our method matches moments for the nuisance parameter
but not for the parameter of interest $\beta$.
This differs from
applications of the expectation consistent framework in which moment matching is pervasive such as in expectation propagation \citep{HernandezLobato2014}.
Implementations of expectation propagation are usually not able to capture dependence among dimensions of the posterior while our method allows for dependence in the $p$-dimensional $\beta$.

Effectively, our method integrates out the nuisance parameter $\eta$ approximately. Integrating out nuisance parameters from the likelihood is not new \citep{Berger1999}, including doing so approximately \citep{Severini2011}. Previous approximations, however, do not apply a data rotation and consider cases where the distribution on the nuisance parameter is regular enough so that a Laplace approximation can be applied. Our method does not need such regularity conditions.

The rotation $Q$ is similar to the projection in the Frisch-Waugh-Lovell theorem \citep[Theorem~11.2.1]{Stachurski2016} for least-squares estimation of a parameter subset.
Our method applies beyond least squares. Also, our estimation of the nuisance parameter through the rotation is merely an intermediate step for inference on $\beta$.
Our method reduces to the algorithm from \citet{vandenBoom2015a} when considering the example in \Sec{}\ref{sec:bvs} with $p = 1$.

\subsection{Estimating $\sigma^2$ and hyperparameters}
\label{sec:sigma}

So far, we have treated $\sigma^2$ as fixed and known. In practice, $\sigma^2$ usually needs to be estimated, as well as any unknown parameters in the prior on $\eta$.
This estimation fits naturally into Step~2 of Algorithm~\ref{alg:irga} as the methods that can be used there frequently come with such estimation procedures: See for instance \Sec{}\ref{sec:vamp_sigma} and \Sec{}\ref{sec:lasso} of the Supplementary Material. The resulting estimates can then be plugged into Step~3. By doing so, only the $(n-p)$-dimensional submodel \eqref{eq:rot2} informs the estimates of these parameters and not the $p$-dimensional submodel \eqref{eq:rot1}.
We expect \eqref{eq:rot2} to contain the vast majority of information on the unknown parameters if $(n-p) \gg p$, which is often the case in scenarios of interest.

\section{Examples of nuisance parameters $\eta$}
\label{sec:examples}

\subsection{Adjusting for high-dimensional covariates}
\label{sec:covariates}

Section~\ref{sec:examples} provides examples of the  general setting  of model \eqref{eq:lm} that demonstrate the utility of the integrated rotated Gaussian approximation in  Algorithm~\ref{alg:irga}.
As a first example, consider $\eta = Z\alpha$ with $Z$ a known $n\times q$ feature matrix and $\alpha$ an unknown $q$-dimensional parameter with $q\gg n$. 
Then, the model in \eqref{eq:lm} becomes
$y \sim \N\left(X \beta + Z \alpha,\, \sigma^2 I_n \right)$,
so that we are adjusting for high-dimensional covariates $Z$ in performing inference on the coefficients $\beta$ on the predictors $X$ of interest.
One way to deal with the fact that the number of covariates $q$ exceeds the number of observations $n$ is by inducing sparsity in $\alpha$ via its prior $\Pi(\alpha)$.
We consider the spike-and-slab prior, $\alpha_j\sim \lambda\, \N(0,\,\psi) + (1-\lambda)\,\delta(0)$
independently for $j=1,\dots,q$,
where $\lambda = \pr(\alpha_j\ne 0)$ is the prior inclusion probability,
$\psi$ the slab variance,
and $\delta(0)$ a point mass at zero.
By specifying $\law(\alpha)$, we have also defined $\law(\eta) = \law(Z\alpha)$.
Since each $\law(\alpha_j)$ is a mixture of a point mass and a Gaussian, $\Pi(\alpha)$ and thus $\Pi(\eta)$ are mixtures of $2^q$ Gaussians.
As a result, computation of $\pi(\beta\mid y)$ in \eqref{eq:postB} involves summing over these $2^q$ components.
This is infeasible for large $q$.

Algorithm~\ref{alg:irga} provides an approximation $\hat{\law}(\beta\mid y)$ while avoiding the exponential computational cost.
Step~2 in Algorithm~\ref{alg:irga} requires choice of an estimation algorithm.
Substituting $\eta = Z\alpha$ into \eqref{eq:rot2} yields
${S}^\T y \sim \N({S}^\T Z \alpha,\, \sigma^2 I_{n-p})$, which is
a linear model with $(n-p)$ observations and design matrix ${S}^\T Z$.
As such, methods for linear regression with spike-and-slab priors can produce an approximation to $\Pi(\alpha\mid{S}^\T y)$
and thus the estimates $\hat{\mu}$ and $\hat{\Sigma}$ in \eqref{eq:gauss}.
We choose vector approximate message passing \citep{Rangan2017}, detailed in \Sec{}\ref{sec:vamp} of the Supplementary Material,
to approximate $\Pi(\alpha\mid{S}^\T y)$
because of its computational scalability and accuracy.
The computational scalability limits the size of $q$.
For instance, \Sec{}\ref{sec:snp} considers a subset of $q=10,000$ SNPs as using all SNPs was computationally infeasible.
As a more scalable alternative, we consider
the debiased lasso \citep{Javanmard2013} in \Sec{}\ref{sec:diabetes} as it can also approximate $\Pi(\alpha\mid{S}^\T y)$ as detailed in \Sec{}\ref{sec:lasso} of the Supplementary Material.
A $q$ in the millions is feasible with embarrassingly parallel split-and-merge strategies \citep{Song2014}.
The $q$-dimensional distribution
$\Pi(\alpha\mid {S}^\T y)$
is possibly highly non-Gaussian, being a mixture of Gaussians.
At the same time,
the $p$-dimensional distribution
${\Pi({M}^\T Z\alpha\mid{S}^\T y)} = {\Pi({M}^\T \eta\mid{S}^\T y)}$
can be nearly Gaussian
such that the approximation in \eqref{eq:gauss} is accurate
as discussed in \Sec{}\ref{sec:galen}.

\subsection{Bayesian variable selection}
\label{sec:bvs}

For a second application of \eqref{eq:lm}, consider the linear model
$y\sim\N(A\theta,\, \sigma^2 I_n)$
where $A$ is a known $n\times r$ design matrix and $\theta$ an unknown $r$-dimensional parameter.
Variable selection is the problem of determining which entries of $\theta$ are non-zero.
Modeling the data in a Bayesian fashion provides a natural framework to evaluate statistical evidence via the posterior $\law(\theta\mid y)$.
A standard variable selection prior $\law(\theta)$ is the spike-and-slab prior defined by
$\theta_j\sim \lambda\, \N(0,\,\psi) + (1-\lambda)\,\delta(0)$
independently for $j=1,\dots,p$.
As in \Sec{}\ref{sec:covariates}, the cost of computing the exact posterior with a spike-and-slab prior grows exponentially in $r$. Therefore, computation of $\Pi(\theta\mid y)$ is infeasible for $r$ beyond moderate size.
A variety of approximation methods exist for larger $r$ including
Monte Carlo \citep{George1993,George1997,Ohara2009}, variational Bayes \citep{Carbonetto2012,Ormerod2017}, and expectation propagation \citep{HernandezLobato2014}.

Monte Carlo methods do not scale well with the number of predictors $r$.
For $r$ even moderately large, the $2^r$ possible non-zero subsets of $\theta$ is so huge that there is no hope of visiting more than a vanishingly small proportion of models.
The result is high Monte Carlo error in estimating posterior probabilities, with almost all models assigned zero probability as they are never visited.
As an alternative to Monte Carlo sampling,
fast approximation approaches for Bayesian variable selection include variational Bayes \citep{Carbonetto2012,Ormerod2017} and expectation propagation \citep{HernandezLobato2014}. Their accuracy, however, does not come with theory guarantees.
Our method, which applies to variable selection as detailed in the next paragraph, allows for theoretical analysis as \Sec{}\ref{sec:theory} shows.

In variable selection, often the main question asked is whether $\theta_j\ne 0\ (j = 1,\dots,r)$ as measured by the posterior inclusion probability ${\pr(\theta_j\ne 0\mid y)}$.
Algorithm~\ref{alg:irga} can estimate ${\pr(\theta_j\ne 0\mid y)}$:
Let $p<r$ elements from $\theta$ constitute $\beta$ and let the other $q = r-p$ elements in $\theta$ constitute $\alpha$.
Then, $A\theta = X\beta + Z\alpha$ where $X$ and $Z$ consist of the respective columns in $A$, and $\law(\alpha,\beta)=\law(\alpha)\,\law(\beta)$ since $\law(\theta)= \prod_{j=1}^r \law(\theta_j)$.
This set-up is the same as in \Sec{}\ref{sec:covariates} and
Algorithm~\ref{alg:irga} approximates $\law(\beta\mid y)$ as in \Sec{}\ref{sec:covariates}.
Assuming $\theta_j$ is contained in $\beta$,
an approximation of $\law(\theta_j\mid y)$
can be obtained as a marginal distribution of $\hat{\law}(\beta\mid y)$.
Repeating Algorithm~\ref{alg:irga} with different splits of $\theta$ into $\beta$ and $\alpha$ provides estimates of all $\pr(\theta_j\ne 0\mid y)\ (j = 1,\dots,r)$.
Computations for these different splits can run in parallel.

The approximation accuracy is not very sensitive to how $\theta$ is split into $\alpha$ and $\beta$, and to $p$ per \Sec{}\ref{sec:theta_split} of the Supplementary Material. We therefore use simple sequential splitting, where the first $p$ elements of $\theta$ constitute $\beta$ in the first split, and recommend choosing $p$ based on computational complexity.
Assume that the number of CPU cores is less than the number of variables $r$.
Then, computation time to obtain all $\hat{\pr}(\theta_j\ne 0\mid y)$ is a trade-off between the length $p$ of $\beta$, which affects the cost of each execution of Algorithm~\ref{alg:irga}, and the number $r/p$ of executions of Algorithm~\ref{alg:irga}.
The order of $r$ is limited by the order of $q$, which is again limited by the algorithm chosen for Step~2 of Algorithm~\ref{alg:irga} as discussed in \Sec{}\ref{sec:covariates}.
The complexity in terms of $p$ and $r$ of computing all $\hat{\pr}(\theta_j\ne 0\mid y)$ is $O(r^2 \log^2 r)$ if $p=O(\log r)$ and vector approximate message passing is used as detailed in the next paragraph.

Step~1 of Algorithm~\ref{alg:irga} is the QR decomposition of an $n\times p$ matrix which has complexity $O(np^2)$ \citep[\Sec{}5.2]{Golub1996}.
Step~2 involves vector approximate message passing on $n-p$ observations and $q$ parameters, which has a complexity of $O\{{(n-p+K)}\, q\, {\min(n-p,q)}\}$ where $K$ is the number of message passing iterations as detailed in \Sec{}\ref{sec:vamp_comp} of the Supplementary Material.
Additionally for Step~2, computation of $S^\T y$ and $S^\T Z$, which are the observations and design matrix in \eqref{eq:rot2}, and computing $\hat{\mu}$ and $\hat{\Sigma}$ in \eqref{eq:gauss} from the message passing output is $O(n^2 q)$.
Computing Step~3
with the spike-and-slab prior $\law(\beta)$ is $O(2^p \, p^3)$, ignoring dependence on $n$.
The complexity of obtaining all ${\hat{\pr}(\theta_j\ne 0\mid y)}$ by applying Algorithm~\ref{alg:irga} $r/p$ times is thus
${O\{ (r/p) (q + 2^p\, p^3 ) \}} = {O\{ (r/p) (r-p + 2^p\, p^3) \}}$, ignoring dependence on $n$ and $K$.
For $p = O(\log r)$, this complexity reduces to $O(r^2 \log^2 r)$.

\subsection{Non-parametric adjustment for covariates}
\label{sec:nonparam_covariates}

As a last example, let $\eta_i = (g\circ f)(z_i)\ (i=1,\dots,n)$ where $g:\mathbb{R}\to\mathbb{R}$ is a known, differentiable, non-linear function,
$f:\mathbb{R}^q\to\mathbb{R}$ is an unknown function,
$g\circ f:\mathbb{R}^q\to\mathbb{R}$ is $g$ composed with $f$,
and $z_i$ is a $q$-dimensional feature vector.
Then, $\eta_i$ provides a non-parametric adjustment for the covariate $z_i$ in performing inferences on the effect of $x_i$ on $y_i$.
Take $f$'s prior as a Gaussian process that induces a prior $\law(\eta)$.
Algorithm~\ref{alg:irga} applies if a Gaussian approximation ${\hat{\law}({M}^\T \eta \mid {S}^\T y)}$ is available:
Submodel \eqref{eq:rot2} reduces to
${S}^\T y \sim \N\{{S}^\T G(F),\, \sigma^2 I_{n-p} \}$ where
$F = \{f(z_1),\dots,f(z_n)\}^\T$
and
$G(F) = \{g(F_1),\dots,g(F_n)\}^\T$,
which is a non-linear Gaussian model as studied in \citet{Steinberg2014}.
Linearizing $G$ using a first-order Taylor series yields a Gauss-Newton algorithm for a Laplace approximation
of ${\law(F \mid {S}^\T y)}$ as detailed in \Sec{}\ref{sec:gauss_newton} of the Supplementary Material.
Based on that approximation, compute $\hat{\mu}$ and $\hat{\Sigma}$ in \eqref{eq:gauss}, for instance by sampling
$F$ from a Laplace approximation ${\hat{\law}(F \mid {S}^\T y)}$
and computing the sample mean and covariance of ${M}^\T G(F)$ since ${M}^\T \eta = {M}^\T G(F)$.

\section{Analysis of integrated rotated Gaussian approximation}
\label{sec:theory}

\subsection{Approximation accuracy}
\label{sec:divBound}
This section provides theoretical guarantees on the accuracy of our posterior approximation framework. We begin with a general upper bound  in terms of the accuracy of the approximation for the projected nuisance parameter. 
For this, denote the distribution of the $p$-dimensional $a+b$ where $b\sim\N(0,\, \sigma^2 I_p)$ by $\law(a) \ast \N_{\sigma^2}$.
Define the Kullback-Leibler divergence from $\law(b)$ to $\law(a)$ as $D\{\law(a)\ \|\ \law(b)\} = \int \log\{ \pi(a) / \pi(b) \}
	d\Pi(a)$.

At a high level, it is clear that the accuracy of  the approximation $\hat{\law}(\beta \mid y)$ defined in \eqref{eq:ira}  depends on the accuracy of the approximation   $\hat{\law}\left( {M}^\T \eta \mid {S}^\T y \right)$. The following result  quantifies the nature of this dependence in the setting where the data are generated from the prior predictive distribution. This result applies generally for any approximation $\hat{\law}\left( {M}^\T \eta \mid {S}^\T y \right)$ and thus includes the Gaussian approximation \eqref{eq:gauss} used in Algorithm~\ref{alg:irga} as a special case.

\begin{theorem} \label{thm:divBoundBayes}
Let $y$ be distributed according to the  model in \eqref{eq:lm} with $\beta\sim \Pi(\beta)$ and $\eta\sim\law(\eta)$ distributed according to their priors. Conditional on any realization $S^\T y$, the posterior approximation $\hat{\law}(\beta\mid y)$ described in  \eqref{eq:ira} satisfies
\[
	\ee\left[ D\left\{\law(\beta\mid y)\ \|\ \hat{\law}(\beta\mid y)\right\} \ \middle\vert\ {S}^\T y \right] 
	\le D \left\{ \law( M^\T \eta \mid S^\T y)  \ast \N_{\sigma^2} \ \|\ \hat{\law}( M^\T \eta \mid S^\T y) \ast \N_{\sigma^2}   \right\},
\]
where the expectation on the left is with respect to the conditional distribution of $y$ given $S^\T y$. 
\end{theorem}

A particularly useful property of Theorem~\ref{thm:divBoundBayes} is that the upper bound does not depend in any way on the prior $\Pi(\beta)$. This differs from some of the related work on posterior approximation, such as \citet{Huggins2017}, which requires additional smoothness constraints, and thus  excludes certain priors such as the spike-and-slab prior in \Sec{}\ref{sec:covariates}. Another useful property of Theorem~\ref{thm:divBoundBayes} is that it does not require any assumptions about the extent to which the exact posterior ${\law( M^\T \eta \mid S^\T y)}$ is concentrated about the ground truth. As a consequence, this result is relevant for non-asymptotic settings where there may be high uncertainty about $\eta$.

\subsection{Accuracy of the Gaussian approximation}
\label{sec:galen}

Next, we provide theoretical justification for a Gaussian approximation to $\law\left( {M}^\T \eta \mid {S}^\T y \right)$ by showing that such an approximation can be accurate even when the prior on $\eta$ is highly non-Gaussian. Without loss of generality, we focus on the set-up of \Sec{}\ref{sec:covariates} where the nuisance term has the form  $\eta = Z\alpha$ with a known $n \times q$ feature matrix $Z$ and unknown parameter vector $\alpha$. In this setting, the projected nuisance parameter $M^\T \eta$ can be expressed as $M^\T Z \alpha$ where $M^\T Z$ is a $p \times q$ matrix with $p \ll q$. There are no constraints on the dimension $n$ other than $n \geq p$. 

As motivation for a Gaussian approximation to the projected nuisance term, consider the special case where the conditional distribution $\law( \alpha \mid S^\T y)$ is a product measure with uniformly bounded second moments. Under regularity assumptions on the columns of $M^\T Z$, the multivariate central limit theorem  combined with the assumption $p \ll q$ implies that the distribution of the projection $M^\T Z \alpha$ is close to the Gaussian distribution with the same mean and covariance. By contrast, the unprojected $n$-dimensional nuisance term $\eta = Z \alpha$ can be very far from Gaussian, particularly if $n$ is of a similar order to $q$.

More realistically, one may envision settings where the entries of $\law( \alpha \mid S^\T y)$  are not independent but are weakly correlated on average. In this case, the usual central limit theorem does not hold because one can construct counterexamples in which the normalized sum of dependent but uncorrelated variables is far from Gaussian. Nevertheless, a  classic result due to \citet{Diaconis1984} suggests that these counterexamples are atypical. Specifically, if one considers a weighted linear combination of the entries in $\alpha$, then approximate Gaussianity holds for most choices of the weights, where most is quantified with respect to the uniform measure on the sphere. The implications of this phenomenon have been studied extensively in the context of statistical inference \citep{Hall1993, Leeb2013}, and \citet{Meckes2012} and \citet{Reeves2017} provide approximation bounds for the setting of multidimensional linear projections.

In the context of our approximation framework, these results imply that a Gaussian approximation is accurate for most, but not necessarily all,  instances of the $p \times q$ feature matrix $M^\T Z$.  To make this statement mathematically precise, we consider the expected behavior when the rows of $Z$ are drawn independently from the $q$-dimensional Gaussian distribution $\N(0,\, \Lambda)$ where $\Lambda$ is positive definite.  As in the rest of the paper, we assume that $X$ is fixed and arbitrary.  Under these assumptions, the rows of the projected matrices $M^\T Z$ and $S^\T Z$ are independent with the same distribution as in $Z$. 

Our results depend on certain properties of the conditional distribution $\law( \alpha \mid S^\T y, S^\T Z)$.  Let $\xi$ and $\Psi$ denote the mean and covariance of $\law( \alpha \mid S^\T y, S^\T Z)$, respectively. Define 
\[
m_1 = \ee\left\{  \left |  \frac{ \| \Lambda^{\frac{1}{2}}( \alpha -  \xi) \|^2}{  \gtr(\Lambda \Psi)}  - 1 \right| \, \bigm\vert S^\T y,\, S^\T Z\right\},
\qquad
m_2   = \frac{ \gtr\{ (\Lambda \Psi)^2\}}{ \gtr(\Lambda \Psi)^2}.
\]
The term $m_1$  provides a measure of the concentration of $\| \Lambda^{1/2} (\alpha - \xi)\|^2$ about its mean and satisfies  $0 \le m_1 \le 2$. The term $m_2$ provides a measure of the average correlation between the entries of $\Lambda^{1/2} \alpha$ and satisfies  $1/q \le m_2 \le 1$ with equality on the left when $\Lambda \Psi$ is proportional to the identity matrix and equality on the right when $\Lambda  \Psi$ has rank one. 

Given estimates $\hat{\xi}$ and $\hat{\Psi}$ that are functions of $S^\T y$ and $S^\T Z$, we consider the Gaussian approximation
\begin{align}
\hat{\law}( M^\T \eta \mid S^\T y, S^\T Z) = \N\{ M^\T Z \hat{\xi},\, \gtr( \Lambda \hat{\Psi})  I_p\}. \label{eq:eta_approx_gamma_hat}
\end{align}
The covariance is chosen independently of $M^\T Z$ and depends only on a scalar summary of the estimated covariance. The following result bounds the accuracy of this approximation in terms of the terms $m_1$ and $m_2$ and the accuracy of the estimated mean and covariance.

\begin{theorem}\label{thm:CCLT_bound} 
Conditional on any $S^\T y$ and $S^\T Z$, the Gaussian approximation in \eqref{eq:eta_approx_gamma_hat} satisfies
 \begin{align*}
 \ee_{M^\T Z}\left[D \left\{ \law( M^\T \eta \mid S^\T y, S^\T Z)  \ast \N_{\sigma^2} \ \|\ \hat{\law}( M^\T \eta \mid S^\T y, S^\T Z) \ast \N_{\sigma^2}   \right\} \right]  \le \delta_1 + \delta_2,
\end{align*}
where the expectation is with respect to $M^\T Z$ and 
\begin{align*}
\delta_1 & = 3 p\, \left[ m_1\, \log\left\{1 + \frac{\gtr(\Lambda \Psi)}{\sigma^2 }\right\}   + m_2^{\frac{1}{4}}   + m_2^{\frac{1}{2}}\,  \left\{ 1 + \frac{ 3  \gtr(\Lambda \Psi) }{\sigma^2} \right\}^{\frac{p}{4}}   \right], \\
\delta_2 & = \frac{ p\,  \| \Lambda^{\frac{1}{2}} (\xi - \hat{\xi})\|^2}{ 2 \sigma^2}  +  \frac{p}{ 2 \sigma^2}\, \left\{  \gtr(\Lambda \Psi)^{\frac{1}{2}} - \gtr(\Lambda  \hat{\Psi})^{\frac{1}{2}} \right\}^2.
\end{align*}
\end{theorem}

This result is meaningful when $p \ll q$  and the noise variance is non-negligible compared to the covariance of the nuisance term such the ratio $\gtr(\Lambda \Psi)/\sigma^2$ is bounded from above. Then, $\delta_1$ converges to zero as $m_1$ and $m_2$ become small.  The term $\delta_2$ quantifies the effect of mismatch between the first and second moments of  $\law( \alpha \mid S^\T y, S^\T Z)$ and their approximations. The dependence on the second moments appears only in the terms $ \gtr(\Lambda \Psi)$ and $ \gtr(\Lambda  \hat{\Psi})$. Thus, this bound can be small even if the approximation $\hat{\Psi}$ is very different from the true covariance $\Psi$.

To illustrate the significance of our results, consider two scaling regimes. First, if $n \ll q$ then the same arguments used in the proof of Theorem~\ref{thm:CCLT_bound} can be used to show that the distribution of the $n$-dimensional nuisance term $\eta$ is also approximately Gaussian. Then, our approximation framework is well motivated, but does not differ fundamentally from existing approaches that apply a Laplace approximation directly on the unrotated data. The second, and more interesting, regime occurs when $n \approx q$ or $n \gg q$. Then, the $n$-dimensional nuisance term is non-Gaussian in general, because there exists a near isometry between $\eta$ and $\alpha$.  Our approximation framework can provide significant gains by taking this non-Gaussianity into account when estimating the mean and covariance. Moreover, combining Theorems~\ref{thm:divBoundBayes} and \ref{thm:CCLT_bound} provides an upper bound on the error of the approximation to the posterior of  $\beta$ described in Algorithm~\ref{alg:irga}.  In particular, if  the approximations of the mean and covariance are accurate enough, then this approximation error converges to zero as the terms $m_1$ and $m_2$ become small.

\subsection{Variable selection consistency}
\label{sec:consistency}

Finally, we provide guarantees for variable selection consistency of \eqref{eq:irga}, which only considers $\beta$ in contrast to \Sec{}\ref{sec:bvs}.
Let the set $\gamma\subset\{1,\dots,p\}$ contain all indices $j$ such that $\beta_j \ne 0$.
Define $\gamma^0$ analogously for a non-random $\beta^0$.
Variable selection consistency as in \citet{Fernandez2001} and \citet{Liang2008} means that, for $y \sim \N( X \beta^0 + \eta^0,\, \sigma^2 I_n)$, the posterior probability of the true model $\gamma^0$ converges to one, ${\pr(\gamma = \gamma^0 \mid y)}\to 1$ as $n\to\infty$ where $p$ does not change with $n$.
It is desirable for a posterior approximation to inherit this property.
Monte Carlo approximations do, but only if they are run for an infinite amount of time.
Our approximation bypasses the need for such sampling,
instead requiring mean and variance estimation for \eqref{eq:gauss},
while inheriting the consistency property
if $\hat{\law}\left( {M}^\T \eta \mid {S}^\T y \right)$ concentrates appropriately.
Relatedly, \citet{Ormerod2017} established such consistency for their variational Bayes algorithm. More recently, K.\ Ray and B.\ Szab\'o (arXiv:1904.07150) showed optimal convergence rates of variable selection using variational Bayes with different priors
than we consider here.

Let the $|\gamma|$-dimensional vector $\beta_\gamma$ consist
of the elements in $\beta$ with indices in $\gamma$,
and
the $n\times |\gamma|$ matrix $X_\gamma$ consist of the columns of $X$ with indices in $\gamma$.
Then, specifying $\Pi(\gamma)$ and $\Pi(\beta_{\gamma}\mid\gamma)$ defines $\Pi(\beta)$.
We consider $g$-priors \citep{Zellner1986}:
\begin{equation} \label{eq:gprior}
	\beta_\gamma \mid \gamma \sim N\left\{0,\ \sigma^2 g_n \left( X_\gamma^\T X_\gamma \right)^{-1} \right\},
	\qquad g_n\in(0,\, \infty).
\end{equation}
\citet{Liang2008} showed variable selection consistency
for priors of this form.
Our approximation inherits this property under the additional assumption \eqref{eq:eta_assum} on $\hat{\law}\left( {M}^\T \eta \mid {S}^\T y \right)$ and $g_n$.
This is an assumption on $g_n$ and $\sigma^2$ jointly since
$\hat{\law}\left( {M}^\T \eta \mid {S}^\T y \right)$
depends on $\sigma^2$.
Otherwise, the sensitivity on $\sigma^2$ is limited since the property considers the asymptotic regime $n\to\infty$,
when the signal-to-noise ratio goes to infinity regardless of $\sigma^2$.

\begin{theorem} \label{thm:consistencyS}
Let $\Pi(\beta_\gamma)$ be the $g$-prior on $\beta_\gamma$ from \eqref{eq:gprior}. Assume that $g_n$ in \eqref{eq:gprior}, $\Pi(\gamma)$, and $X$ satisfy
$\pr(\gamma = \gamma^0) > 0$,
$\lim_{n\to\infty} \| \{I_n -  X_{\gamma} (X_{\gamma}^\T X_{\gamma})^{-1} X_{\gamma}^\T \} X\beta^0 \|/n > 0$
for any $\gamma$ not containing $\gamma^0$,
$g_n \to \infty$,
and $\log (g_n) / n \to 0$,
which are standard assumptions used in \citet{Fernandez2001} and \citet{Liang2008} as detailed in \Sec{}\ref{sec:consistency_proof} of the Supplementary Material.
Let $y$ be distributed according to the data-generating model in \eqref{eq:lm} with $\beta$ and $\eta$ fixed to $\beta^0$ and $\eta^0$, respectively. 
Assume that $\hat{\law}\left( {M}^\T \eta \mid {S}^\T y \right)$ concentrates appropriately in that
\begin{equation} \label{eq:eta_assum}
\frac{\| M^\T \eta - M^\T \eta^0 \|^2}{\log g_n} \to 0,
\end{equation}
in probability with respect to
$M^\T \eta \sim
\hat{\law}\left( {M}^\T \eta \mid {S}^\T y \right)$ and $y$.
Let $\hat{\law}(\beta\mid y)$ be as in \eqref{eq:ira}.
Then, $\hat{\pr}(\gamma = \gamma^0 \mid y ) \to 1$ in probability with respect to $y$ as $n\to\infty$.
\end{theorem}

\section{Simulation studies and applications}

\subsection{Non-parametric adjustment for covariates}
\label{sec:GPsimul}

\begin{figure}[tbp]
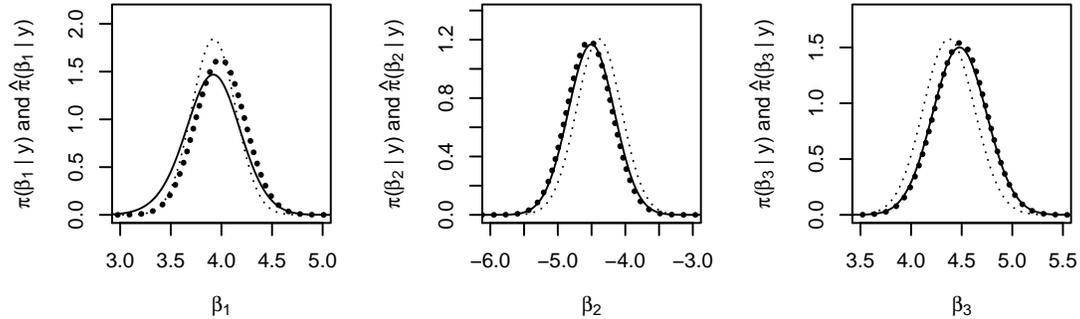

	\centering
	\figurebox{}{\textwidth}{}[simulation_nonparametric.eps]
	\caption{Marginal posterior density estimates from the simulation in \Sec{}\ref{sec:GPsimul} with the solid line representing the Gibbs estimate $\pi(\beta_j\mid y)$, the thick dotted line the estimate $\hat{\pi}(\beta_j\mid y)$ from Algorithm~\ref{alg:irga}, and the thin dotted line the estimate resulting from ignoring the nuisance parameter.}
	\label{fig:GPsimul}
\end{figure}

Consider the set-up from \Sec{}\ref{sec:nonparam_covariates}
with $g(a) = a^2$ and $q=1$.
We assign $f:\mathbb{R}\to\mathbb{R}$
a zero-mean Gaussian process prior with a squared exponential covariance function such that $\cov\{f(z_i),\, f(z_j)\} = \exp\{ -(z_i-z_j)^2/10 \}\ (i,j=1,\ldots,n)$,
and
$\beta\sim\N(0, 16I_p)$.
Set $n=100$, $p=3$, and $\sigma^2 = 1$. We draw the rows of $X$ independently from $\N(0_{p\times 1}, \Phi)$ 
where $\Phi$ is a Toeplitz matrix defined so that its first row equals $(0.9^0, \ldots, 0.9^p)$. Then, the columns of $X$ are correlated.
The features $z_i\ (i=1,\dots,n)$ equal the $i$th element of the first column of $X$.
Generate $y$ according to \eqref{eq:lm} with $f$ equal to a draw from its prior distribution and $\beta=(4,-4,4)^\T$.

We approximate the posterior $\law(\beta\mid y)$ using a random walk Metropolis-Hastings
algorithm on $f$
with 10,000 burnin and 90,000
recorded iterations.
We marginalize out $\beta$
since $\law(\beta\mid f, y)$ is analytically available, allowing
approximation of $\pi(\beta\mid y)$ with samples from $\law(f\mid y)$.
Algorithm~\ref{alg:irga} also provides $\hat{\law}(\beta\mid y)$
per \Sec{}\ref{sec:nonparam_covariates}.
Lastly, ignoring the non-parametric nuisance parameter by setting $\eta_i = (g\circ f)(z_i) = 0$
yields a simpler approximation.
The Metropolis-Hastings algorithm took 6 minutes while our method finished in 2 seconds.
The resulting posterior density estimates for $\beta_j\ (j=1,\ldots,p)$ are in Fig.~\ref{fig:GPsimul}.
Taking the Metropolis-Hastings estimate as the gold standard,
our method yields an approximation that matches the location and spread of the posterior better than the result from ignoring the non-parametric nuisance term $\eta$.

\subsection{Bayesian variable selection}
\label{sec:diabetes}

We consider the diabetes data from \citet{Efron2004}
as it is a popular example of variable selection
with collinear predictors
\citep{Park2008,Polson2013}.
The outcome $y$ measures disease progression one year after baseline for $n=442$ patients with diabetes. The $r=64$ predictors come from 10 covariates with their squares and interactions.
The outcome and predictors are standardized to have zero mean and unit norm.
Consider the variable selection set-up from \Sec{}\ref{sec:bvs}
with prior inclusion probability $\lambda=1/2$ and $\psi=1$.
Usually, one would not use scalable approximations for such a moderate-dimensional problem as a Gibbs sampler can provide accurate estimates.
The latter is why we include it here as these accurate estimates enable assessment of the approximation accuracy of scalable methods.

We
estimate the posterior inclusion probabilities $\pr(\theta_j \ne 0 \mid y)\ (j=1,\ldots, r)$ using 1) a Gibbs sampler with 10,000 burnin and 90,000
recorded iterations, Algorithm~\ref{alg:irga} as described in \Sec{}\ref{sec:bvs} using 2) vector approximate message passing and 3) the debiased lasso in Step~2 with $p=4$ as suggested by $p = O(\log r)$ and parallelization across 8 CPU cores, 4) expectation propagation as in \citet{HernandezLobato2014}, and 5) variational Bayes as in \citet{Carbonetto2012}.
To implement expectation propagation and variational Bayes, we used
the R code from \url{https://jmhl.org/publications/} dated January 2010
and
the R package `varbvs' version 2.5-7, respectively.
Results from the variational Bayes algorithm by \citet{Ormerod2017} are omitted as the method from \citet{Carbonetto2012} outperforms it in the scenarios that we consider.
Since the error variance is unknown, we assign it the prior
$1/\sigma^2\sim\Ga(1,1)$, a gamma distribution with unit shape and rate parameter.
The Gibbs sampler
incorporates this prior.
Algorithm~\ref{alg:irga} estimates $\sigma^2$ as described in \Sec{}\ref{sec:sigma}, and \Sec{}\ref{sec:vamp_sigma} and \Sec{}\ref{sec:lasso} of the Supplementary Material.
Expectation propagation estimates $\sigma^2$ by maximizing approximate evidence \citep{HernandezLobato2014}.
The R package `varbvs' \citep{Carbonetto2012} uses approximate maximum likelihood for $\sigma^2$ within the variational Bayes method.

As discussed in \Sec{}\ref{sec:bvs}, determining whether posterior inclusion probabilities from a Gibbs sampler are accurate is non-trivial.
Overlapping batch means \citep[\Sec{}3]{Flegal2010} estimates their average Monte Carlo standard error as
0.0015 in this application.

\begin{table}[tbp]
\caption{Summary statistics of the absolute difference between the Gibbs sampler estimates and the approximations of the posterior log odds of inclusion for the application in \Sec{}\ref{sec:diabetes}
with computation times.
IRGA and VAMP stand for integrated rotated Gaussian approximation and vector approximate message passing, respectively.
\label{tab:diabetes}}
\centering
\begin{tabular}{r|cccccc|c}
Method & Min & Q1 & Median & Q3 & Max & Mean & \thead{Computation \\ time (seconds)} \\
\hline
IRGA with VAMP & 0.003 & 0.036 & 0.076 & 0.133 & 10.7 & 0.599 & 4.1 \\
IRGA with the debiased lasso & 0.003 & 0.100 & 0.142 & 0.199 & 7.85 & 0.470 & 3.8 \\
Expectation propagation & 0.003 & 0.061 & 0.109 & 0.168 & 11.9 & 0.666 & 0.8 \\
Variational Bayes & 0.002 & 0.093 & 0.124 & 0.166 & 11.6 & 0.667 & 1.0
\end{tabular}
\end{table}

Table~\ref{tab:diabetes} focuses on the errors in the posterior inclusion probability estimates.
An approximation error of $0.01$ is worse when the inclusion probability is $0.01$ versus $0.5$. We therefore transform the probabilities to log odds.
Our method with vector approximate message passing outperforms expectation propagation and variational Bayes as its error is lowest in Table~\ref{tab:diabetes}, though at a higher computational cost.
Our method is slowest but still considerably faster than the Gibbs sampler which took 11 minutes to run.
Since the debiased lasso yielded the worst approximation, we do not consider it in the remainder of this article.

\subsection{Controlling for single-nucleotide polymorphisms}
\label{sec:snp}

The Geuvadis dataset from \citet{Lappalainen2013}, available at \url{https://www.ebi.ac.uk/Tools/geuvadis-das}, contains gene expression data from lymphoblastoid cell lines of $n=462$ individuals from the 1000 Genomes Project along with roughly 38 million SNPs.
We focus on the gene E2F2, ensemble ID ENSG00000007968, as it plays a key role in the cell cycle \citep{Attwooll2004}.
Our focus is on assessing whether expression differs between populations, even after adjusting for genetic variation between individuals.
Specifically, we compare people from British descent with the four other populations given in Table~\ref{tab:snp}.
If such differences occur, they can be presumed to be due to environmental factors that differ between these populations and that relate to E2F2 expression.
We therefore consider the set-up from \Sec{}\ref{sec:covariates} with $y$ the E2F2 gene expressions,
$X$ demographic factors,
and $Z$ containing SNPs we would like to control for.

The demographics in $X$ are gender and the 4 populations with British as the reference group. The matrix $X$ thus has $p=5$ columns.
The covariates $Z$ consist of $q=10,000$ SNPs selected using
sure independence screening \citep{Fan2008}
as vector approximate message passing on all 38 million SNPs was infeasible.
We standardize $y$ and the columns of $X$ and $Z$ to have zero mean and unit variance.
To complete the set-up from \Sec{}\ref{sec:covariates}, set
$\lambda = n/(10\, q)$ and $\psi = 1/n$ for the spike-and-slab prior on $\alpha$
while $\law(\beta)$ is a spike-and-slab with prior inclusion probability $1/2$ and slab variance $1$ such that, a priori, the SNPs do not capture more variation in the outcome than the demographic factors.  This may provide a reasonable default for SNP data, but in other settings, hyperparameter values should be reconsidered.
Vector approximate message passing estimates $\sigma^2$ using the prior $1/\sigma^2\sim\Ga(1,1)$ and employs damping to achieve convergence in this application, as described in \Sec{}\ref{sec:vamp_sigma} and \Sec{}\ref{sec:dampening} of the Supplementary Material, respectively.

\begin{table}[tbp]
\caption{Posterior inclusion probabilities for the demographic factors from the application in \Sec{}\ref{sec:snp}.
IRGA stands for integrated rotated Gaussian approximation.
\label{tab:snp}}
\centering
\begin{tabular}{r|c|cccc}
& & \multicolumn{4}{c}{Population} \\
Method & Gender & Utahn of European ancestry & Finnish & Tuscan & Yoruba \\
\hline
IRGA & 0.83 & 0.96 & 0.96 & 0.92 & 0.00 \\
Ignoring the SNPs & 0.73 & 0.07 & 0.04 & 0.20 & 0.49
\end{tabular}
\end{table}

Table~\ref{tab:snp} contains the resulting posterior inclusion probabilities for the demographic factors, also when not controlling for the SNPs.
The results vary hugely by whether SNPs are controlled for,
with more evidence of a difference in the expression of gene E2F2 by population when controlling for SNPs using Algorithm~\ref{alg:irga}.
Section~\ref{sec:snp_supp} of the Supplementary Material contains additional comparisons with other high-dimensional inference methods.

Section~\ref{sec:add_simulations} of the Supplementary Material contains additional simulation studies.
They further show that integrated rotated Gaussian approximation outperforms variational Bayes and either beats or is on par with expectation propagation in terms of approximation accuracy. This improved accuracy comes with increased computational cost for our method in certain scenarios.

\section{Discussion}

Although our focus was Bayesian inference, our method marginalizes out nuisance parameters from the likelihood for $\beta$ as an intermediate step. This approximate likelihood from \eqref{eq:irga} can be useful in frequentist inference.
It is well known that priors used in Bayesian inference correspond to penalties in frequentist inference. One can think of the log prior for the nuisance parameter $\eta$ as a penalty on $\eta$.
An $L_2$ penalty might not be ideal due to the complex or high-dimensional nature of $\eta$.
Instead, one might want to use sparsity-inducing penalties,
such as $L_1$ or the non-convex smoothly clipped absolute deviation,
which come with attractive theoretical properties \citep{Potscher2009} but can be computationally challenging.
Our method obtains the marginal likelihood for $\beta$ with such penalties on $\eta$, resolving the main computational bottleneck for frequentist inference on $\beta$ in the model of interest in \eqref{eq:lm}.

\section*{Acknowledgment}

This work was partially supported by 
the National Institute of Environmental Health Sciences of the U.S. National Institutes of Health,
the Singapore Ministry of Education Academic Research Fund,
and the Laboratory for Analytic Sciences. Any opinions, findings, conclusions, or recommendations expressed in this material are those of the authors and do not necessarily reflect the views of the Laboratory for Analytic Sciences and/or any agency or entity of the United States Government.

\section*{Supplementary material}

Supplementary material available at {\it Biometrika} online includes the proofs for \Sec{}\ref{sec:theory},
a corollary to Theorem~\ref{thm:divBoundBayes}, details of vector approximate message passing and the Laplace approximation for \Sec{}\ref{sec:nonparam_covariates}, and additional simulation studies.
The R code for the numerical results is available at \url{https://github.com/willemvandenboom/IRGA}.

\small{
\bibliographystyle{biometrika}
\bibliography{biomet.bib}
}

\end{document}




\markboth{W. van den Boom, G. Reeves \and D. B. Dunson}{Integrated rotated Gaussian approximation}

\title{Supplementary material for\\
Approximating posteriors with high-dimensional nuisance parameters via integrated rotated Gaussian approximation}

\author{W. VAN DEN BOOM}
\affil{Yale-NUS College, National University of Singapore, 16 College Avenue West \#01-220, Singapore 138527, Singapore\email{willem@yale-nus.edu.sg}}

\author{G. REEVES \and D. B. DUNSON}
\affil{Department of Statistical Science, Duke University, Box 90251, Durham,\\ North Carolina 27708, U.S.A. \email{galen.reeves@duke.edu} \email{dunson@duke.edu}}

\maketitle

\section{Proof of Theorem~\ref{thm:divBoundBayes}}

\begin{lemma}\label{lem:DtoD}
Let $P(a,b)$ and $Q(a,b)$ be probability measures defined on the same space that have the same $a$-marginal, that is, $P(a) = Q(a)$. Then, 
\[
\ee_{P(b)}\left[  D\left\{ P(a \mid b)  \ \| \ Q(a \mid b)   \right\}  \right]
\leq
\ee_{P(a)}\left[  D\left\{ P(b \mid a)  \ \| \ Q(b \mid a)   \right\}  \right].
\]
\end{lemma}

\begin{proof}
Using the chain rule for Kullback-Leibler divergence \citep[Theorem~2.5.3]{Cover2006} two different ways leads to
\begin{align*}
 D\left\{ P(a,b)    \ \| \ Q(a,b) \right\}
& =  \ee_{P(b)}\left[  D\left\{ P(a \mid b)  \ \| \ Q(a \mid b)   \right\}  \right]  + D\left\{ P(b)    \ \| \ Q(b) \right\}\\
&  =  \ee_{P(a)}\left[  D\left\{ P(b \mid a)  \ \| \ Q(b \mid a)   \right\}  \right]  + D\left\{ P(a)    \ \| \ Q(a) \right\}.
\end{align*}
Hence, the desired result follows from the fact that $D\left\{ P(b)    \ \| \ Q(b) \right\}$ is non-negative, and the assumption $P(a)  = Q(a)$ which implies that $D\left\{ P(a)    \ \| \ Q(a) \right\} = 0$. 
\end{proof}

\begin{proof}[of Theorem~\ref{thm:divBoundBayes}]
The distributions ${\law(\beta, M^\T y \mid S^\T y)} = {\law({M}^\T y \mid {S}^\T y,\beta)}\, {\law(\beta \mid {S}^\T y)}$ and
${\hat{\law}(\beta, M^\T y \mid S^\T y)} = {\hat{\law}({M}^\T y \mid {S}^\T y,\beta)}\, {\law(\beta \mid {S}^\T y)}$
have the same $\beta$-marginal ${\law(\beta \mid {S}^\T y)}$. Hence, we can apply  Lemma~\ref{lem:DtoD}
with $P(a,b) = {\law(\beta, M^\T y \mid S^\T y)}$
and $Q(a,b) = {\hat{\law}(\beta, M^\T y \mid S^\T y)}$:
\begin{multline*}
\ee\left[ D\left\{\law(\beta\mid y)\ \|\ \hat{\law}(\beta\mid y)\right\} \ \middle\vert\ {S}^\T y \right] \\
\begin{aligned}
&= \ee_{\law({M}^\T y \mid {S}^\T y)}\left[ D\left\{\law(\beta\mid M^\T y , S^\T y )\ \|\ \hat{\law}(\beta\mid M^\T y , S^\T y)\right\}  \right]\\
&  \le \ee_{\law(\beta \mid S^\T y)}\left[ D\left\{\law({M}^\T y \mid \beta, S^\T y)\ \|\ \hat{\law}({M}^\T y \mid \beta, S^\T y)\right\}  \right],
\end{aligned}
\end{multline*}
Let $\law(a) \ast \law(b)$ denote the distribution of $a+b$.
Then, \eqref{eq:rot1} provides
\begin{align*}
\law( {M}^\T y \mid \beta, S^\T y)
& = \law( {M}^\T \eta  \mid S^\T y) \ast \N( M^\T X\beta,\, \sigma^2 I_p),\\
\hat{\law}({M}^\T y \mid \beta, S^\T y)
& =\hat{\law}( {M}^\T \eta  \mid S^\T y) \ast \N( M^\T X\beta,\, \sigma^2 I_p).
\end{align*}
Combining the last two displays yields
\begin{multline*}
	\ee\left[ D\left\{\law(\beta\mid y)\ \|\ \hat{\law}(\beta\mid y)\right\} \ \middle\vert\ {S}^\T y \right] \leq \\
\ee_{\law(\beta \mid S^\T y)}\left[ D\left\{ \Pi( M^\T \eta \mid {S}^\T y) \ast \N(M^\T X\beta,\, \sigma^2 I_p) \ \|\ \hat{\Pi}( M^\T \eta \mid {S}^\T y) \ast \N(M^\T X\beta,\, \sigma^2 I_p) \right\} \right].
\end{multline*}
Since the Kullback-Leibler divergence is invariant to one-to-one transformations \citep[Corollary~4.1]{Kullback1951},
the Kullback-Leibler divergence is constant with respect to $M^\T X\beta$.
The required result follows from setting $M^\T X\beta$ equal to zero and dropping the expectation in the right-hand side of the last display.
\end{proof}

\section{Corollary to Theorem~\ref{thm:divBoundBayes}}

Theorem~\ref{thm:divBoundBayes} considered how close our approximation $\hat{\law}(\beta\mid y)$ is to the posterior $\law(\beta\mid y)$.
Alternatively, one may be interested in a scenario where the nuisance parameter $\eta$ equals $\eta^0$, and one would like to do inference without interference from the nuisance term using $\law(\beta\mid y,\eta^0)$, even though $\eta^0$ is unknown.

Define the squared quadratic Wasserstein distance between the distributions $\law(a)$ and $\law(b)$ as
$W_2^2\{\law(a),\, \law(b)\} = \inf \ee( \|a-b\|^2 )$ where $\|\cdot\|$ denotes the Euclidean norm and the infimum is over all joint distributions on $(a,b)$ such that $a\sim \law(a)$ and $b\sim \law(b)$.

\begin{lemma}\label{lem:DtoW}
Let $P$ and $Q$ be distributions on $\mathbb{R}^p$. For any $\sigma^2 > 0$, 
\[
D\left\{ P \ast \N(0,\, \sigma^2 I_p )  \ \| \  Q \ast \N(0,\, \sigma^2 I_p )   \right\}  \le \frac{1}{2 \sigma^2}\, W_2^2( P,\, Q).
\]
\end{lemma}

\begin{proof}
Let $\Pi(a,b)$ be any coupling on $\mathbb{R}^p \times \mathbb{R}^p$ satisfying the marginal constraints $\Pi(a) = P(a)$ and $\Pi(b)  = Q(b)$. By  the convexity of Kullback-Leibler divergence \citep[Theorem~2.7.2]{Cover2006}, Jensen's inequality provides
\begin{align*}
D\left\{ P \ast \N(0,\, \sigma^2 I_p )  \ \| \  Q \ast \N(0,\, \sigma^2 I_p )   \right\}   & \le \ee_{\Pi(a,b)}\left[  D\left\{  \N(a ,\, \sigma^2 I_p )]   \ \| \ \N(b ,\, \sigma^2 I_d)   \right\} \right] \\
&  = \frac{1}{2 \sigma^2}\, E_{\Pi(a,b)} \left( \| a  - b  \|^2 \right),
\end{align*}
where the equality follows from inserting the Gaussian densities into the definition of the Kullback-Leibler divergence.
Recalling the definition of the quadratic Wasserstein distance and choosing the infimum over all couplings $\Pi(a,b)$ of $P$ and $Q$ gives the stated result.
\end{proof}

\begin{corollary} \label{cor:divBoundBayes}
Let $\hat{\law}(\beta\mid y)$ be as in \eqref{eq:ira}.
Let $y$ be distributed according to the data-generating model in \eqref{eq:lm} with $\beta\sim \Pi(\beta)$ distributed according to its prior and $\eta$ fixed to $\eta^0$. 
Then,
\[
	\ee\left[ D\left\{\law(\beta\mid y,\eta^0)\ \|\ \hat{\law}(\beta\mid y)\right\} \ \middle\vert\ {S}^\T y \right]
\leq \,\frac{1}{2\sigma^2}\, \ee_{\hat{\law}(M^\T \eta \mid {S}^\T y)}\left( \left\| M^\T \eta^0- M^\T \eta  \right\|^2 \ \middle\vert\ {S}^\T y \right).
\]
In particular, under the Gaussian approximation $\hat{\Pi}( M^\T \eta \mid {S}^\T y)$ from \eqref{eq:gauss},
\[
	\ee\left[ D\left\{\law(\beta\mid y,\eta^0)\ \|\ \hat{\law}(\beta\mid y)\right\} \ \middle\vert\ {S}^\T y \right]
\leq \,\frac{1}{2\sigma^2}\, \left\{ \left\| M^\T \eta^0- \hat{\mu}  \right\|^2  + \gtr( \hat{\Sigma} ) \right\}.
\]
\end{corollary}

\begin{proof}
Evaluating Theorem~\ref{thm:divBoundBayes} with Lemma~\ref{lem:DtoW}, $\Pi(\eta) = \delta(\eta^0)$, a point mass at $\eta^0$, and recalling the definition of the quadratic Wasserstein distance provides the first inequality.
For the second equality, \eqref{eq:gauss} provides $M^\T \eta^0- M^\T \eta \mid {S}^\T y \sim {\mathcal{N}(M^\T \eta^0- \hat{\mu},\, \hat{\Sigma})}$.
Evaluating the right-hand side of the first inequality with this distribution provides the second inequality.
\end{proof}

Corollary~\ref{cor:divBoundBayes}
links two different quantities of interest.
The left-hand side is
the difference between our approximation $\hat{\law}(\beta\mid y)$ and the exact posterior $\law(\beta\mid y,\eta^0)$.
The right-hand side involves the average squared deviation
of the distribution ${\hat{\law}(M^\T \eta \mid {S}^\T y)}$
from $M^\T \eta^0$.
This deviation can be small while the average squared deviation of $\law(\eta \mid {S}^\T y)$ from $\eta^0$ is large:
The $n$-dimensional $\eta$ can have a potentially high-dimensional distribution while the $p$-dimensional term $M^\T \eta$ is a projection onto the low-dimensional column space of $M$.
In Corollary~\ref{cor:divBoundBayes}, $y$ is distributed according to \eqref{eq:lm} with $\beta\sim \Pi(\beta)$ while $\eta$ is fixed to $\eta^0$. That $\beta$ and $\eta$ are treated differently is a result of their different treatment in Algorithm~\ref{alg:irga}.

Consider asymptotic analysis where, for a sequence of instances of \eqref{eq:lm}, $n\to\infty$ and interest is in the properties of $\hat{\law}(\beta\mid y)$ as $n\to\infty$.
If ${\hat{\law}(M^\T \eta \mid {S}^\T y)}$ contracts around the value $M^\T \eta^0$ as $n\to\infty$,
Corollary~\ref{cor:divBoundBayes}
shows that the posterior approximation from our method converges to the posterior $\law(\beta\mid y,\eta^0)$ based on the likelihood from \eqref{eq:lm} with $\eta$ equal to $\eta^0$. 
This convergence is in terms of Kullback-Leibler divergence which bounds
dissimilarity measures
commonly
used in asymptotic analyses of Bayesian posteriors.
For instance, Bernstein-von Mises theorems often use total variation distance
\citep{Bontemps2011} which Pinsker's inequality bounds by the square root of the Kullback-Leibler divergence.
The finite-sample analysis of
Corollary~\ref{cor:divBoundBayes} therefore gives rise
to asymptotic properties of the approximate posterior $\hat{\law}(\beta\mid y)$ if $\ee_{\hat{\law}(M^\T \eta \mid {S}^\T y)}( \left\| M^\T \eta^0- M^\T \eta  \right\|^2 \ \vert\ {S}^\T y ) \to 0$.
Such asymptotic results for $\hat{\law}(\beta\mid y)$ differ from usual Bayesian asymptotics due to the set-up of Corollary~\ref{cor:divBoundBayes}: The data-generating process involves $\beta\sim\Pi(\beta)$ rather than fixing $\beta$ to a value.
By contrast, $\eta$ is fixed to $\eta^0$ in the data-generating process of Corollary~\ref{cor:divBoundBayes} rather than distributed according to its prior.

\section{Proof of Theorem~\ref{thm:CCLT_bound}}

To simplify notation, define $a = q^{1/2}\Lambda^{1/2} (\alpha-\xi)$, $b_a = q^{1/2}\Lambda^{1/2} (\hat{\xi}-\xi)$, and ${H} = q^{-1/2}{M}^\T Z \Lambda^{-1/2}$ such that the entries of ${H}$ are independent with distribution $\N(0,\ 1/q)$ and ${H} a = {M}^\T Z (\alpha - \xi)$.
Also,
\begin{align*}
\Delta &= D ( \law( {H} a \mid S^\T y, S^\T Z)  \ast \N_{\sigma^2} \ \|\  \N[
	{H} b_a, \,
	\{\gtr( \Lambda \hat{\Psi}) + \sigma^2 \} I_p
]   ), \\
\Delta_1 &= D ( \law( {H} a \mid S^\T y, S^\T Z)  \ast \N_{\sigma^2} \ \|\  \N[
	0, \,
	\{\gtr( \Lambda \Psi) + \sigma^2 \} I_p
]   ), \\
\Delta_2 &= D ( \N[
	0, \,
	\{\gtr( \Lambda \Psi) + \sigma^2 \} I_p
] \ \|\  \N[
	{H} b_a, \,
	\{\gtr( \Lambda \hat{\Psi}) + \sigma^2 \} I_p
] ).
\end{align*}
Here, $ \N[
	{H} b_a, \,
	\{\gtr( \Lambda \hat{\Psi}) + \sigma^2 \} I_p
]$ is a shifted version of the Gaussian approximation in \eqref{eq:eta_approx_gamma_hat}.
We will show that $\Delta$ equals the divergence in Theorem~\ref{thm:CCLT_bound}.
$\Delta_1$ is the Kullback-Leibler divergence from the target distribution to the Gaussian approximation evaluated with the true and approximated mean and covariance. $\Delta_2$ depends on the mismatch in the estimates $\hat{\xi}$, captured by $b_a$, and $\hat{\Psi}$.

\begin{lemma} \label{lem:cclt}
Conditional on any $S^\T y$ and $S^\T Z$, $\ee_{{H}}(\Delta_1) \leq \delta_1$ with $\delta_1$ as in Theorem~\ref{thm:CCLT_bound}.
\end{lemma}

\begin{proof}
Since ${\ee({H} a \mid S^\T y, S^\T Z)} = 0$,
${\cov({H} a \mid S^\T y, S^\T Z)} = \ee({H} a a^\T {H}^\T)$, where we drop the condition on $S^\T y$ and $S^\T Z$ for notation convenience.
By the law of total expectation,
$\ee({H} a a^\T {H}^\T) = \ee\{\ee({H} a a^\T {H}^\T\mid {H})\}
= \ee\{{H} \cov(a) {H}^\T\}$.
Inserting the definition of $a$ and recalling $\cov(\alpha) = \Psi$ yields
$\ee\{{H} \cov(a) {H}^\T\} = \ee({H} q \Lambda^{1/2} \Psi \Lambda^{1/2} {H}^\T)$.
Since $\ee({H}_{ij} {H}_{kl})$ equals $1/q$ if $(i,j)=(k,l)$ and $0$ otherwise,
$\ee({H} q \Lambda^{1/2} \Psi \Lambda^{1/2} {H}^\T)
= \gtr(q \Lambda^{1/2} \Psi \Lambda^{1/2}) I_p / q$.
The cyclic property of the trace now provides
${\cov({H} a \mid S^\T y, S^\T Z)} = \gtr(q \Lambda^{1/2} \Psi \Lambda^{1/2}) I_p / q = \gtr(\Lambda\Psi) I_p$.
Thus,
the mean and covariance of both distributions in $\Delta_1$ are matched.
Therefore, Theorem~2 from \citet{Reeves2017} evaluated with $\epsilon =1$ and $C=3$ yields
\begin{multline} \label{eq:cclt}
\ee_{{H}}(\Delta_1)
	\leq 3p\log\left\{
		1 + \frac{\frac{1}{q} \ee(\|a\|^2)}{\sigma^2}
	\right\}\, \frac{\frac{1}{q} \ee\{| \|a\|^2 - \ee(\|a\|^2) |\}}{ \frac{1}{q}\ee(\|a\|^2) } \\
	+ 3p^{\frac{3}{4}} \left\{
		\frac{\frac{1}{q}E(|a^\T a'|)}{\frac{1}{q}\ee(\|a\|^2)}
	\right\}^{\frac{1}{2}}
	+ 3p^{\frac{1}{4}} \left\{
		1 + \frac{
			\frac{3}{q} \ee(\|a\|^2)
		}{\sigma^2}
	\right\}^{\frac{p}{4}} \,
	\frac{\frac{1}{q}E(|a^\T a'|^2)^{\frac{1}{2}}}{\frac{1}{q}\ee(\|a\|^2)},
\end{multline}
where $a'$ is an independent copy of $a$.
The remainder of this proof is simplifying this bound.

Since $\ee(a) = 0$ and $\cov(a) = q\Lambda^{1/2} \Psi \Lambda^{1/2}$,
\[
\begin{aligned}
	q^{-2} E(|a^\T a'|^2)
	&= q^{-2} \ee( a^\T a' a'^\T a )
	= q^{-2} \gtr\{ \ee( a a^\T a' a'^\T ) \} \\
	&= q^{-2} \gtr\{ \cov(a)^2 \}
	= \gtr(\Lambda^{\frac{1}{2}} \Psi \Lambda \Psi \Lambda^{\frac{1}{2}} )
	= \gtr\{ (\Lambda\Psi)^2 \},
\end{aligned}
\]
and
\[
	\frac{1}{q}\ee(\|a\|^2)
	= \frac{1}{q} \gtr\{\cov(a)\}
	= \frac{1}{q} \gtr(q\Lambda^{\frac{1}{2}} \Psi \Lambda^{\frac{1}{2}})
	= \gtr(\Lambda \Psi).
\]
Therefore,
\[
	\frac{\frac{1}{q}\ee\{| \|a\|^2 - \ee(\|a\|^2) |\}}{\frac{1}{q}\ee(\|a\|^2)}
	= \ee\left\{ \left|
		\frac{\|a\|^2}{\ee(\|a\|^2)} - 1
	\right| \right\}
	= \ee\left\{ \left|
		\frac{\|q^{\frac{1}{2}}\Lambda^{\frac{1}{2}} (\alpha-\xi)\|^2}{q \gtr(\Lambda \Psi)} - 1
	\right| \right\}
	= m_1,
\]
and, by Jensen's inequality,
\[
	\left\{
		\frac{\frac{1}{q}E(|a^\T a'|)}{\frac{1}{q}\ee(\|a\|^2)}
	\right\}^2
	\leq
	\left\{
		\frac{\frac{1}{q}E(|a^\T a'|^2)^{\frac{1}{2}}}{\frac{1}{q}\ee(\|a\|^2)}
	\right\}^2
	= \frac{q^{-2} E(|a^\T a'|^2)}{\gtr(\Lambda \Psi)^2}
	= m_2.
\]
Inserting the last three displays and $p^{1/4} \leq p^{3/4} \leq p$ into \eqref{eq:cclt} provides the required result.
\end{proof}

\begin{lemma} \label{lem:KLdiff}
Conditional on any $S^\T y$ and $S^\T Z$, $\ee_{{H}}(\Delta_2) \leq \delta_2$ with $\delta_2$ as in Theorem~\ref{thm:CCLT_bound}.
\end{lemma}

\begin{proof}
Combining \eqref{eq:eta_approx_gamma_hat}, Lemma~\ref{lem:DtoW}, and
the evaluation of the quadratic Wasserstein distance between two Gaussians from \citet{Dowson1982} yields
\begin{equation} \label{eq:dowson}
\begin{aligned}
	\Delta_2 &\leq \frac{1}{2\sigma^2} \left[
		\| {H} b_a \|^2
		+ \gtr \left\{
			\gtr(\Lambda\Psi)I_p + \gtr(\Lambda\hat{\Psi}) I_p
			- 2\gtr(\Lambda\hat{\Psi})^{\frac{1}{2}}\gtr(\Lambda\Psi)^{\frac{1}{2}} I_p
		\right\}
	\right] \\
	&= \frac{1}{2\sigma^2} \left[
		\| {H} b_a \|^2
		+ p \{
			(\Lambda\Psi)^{\frac{1}{2}} - (\Lambda\hat{\Psi})^{\frac{1}{2}}
		\}^2
	\right].
\end{aligned}
\end{equation}
Recalling $b_a = q^{1/2}\Lambda^{1/2} (\hat{\xi}-\xi)$,
\[
\begin{aligned}
	\ee_{{H}} \left( \| {H} b_a \|^2 \right)
	&= \ee_{{H}} \left\{ \| q^{\frac{1}{2}}{H} \Lambda^{\frac{1}{2}} (\xi - \hat{\xi}) \|^2 \right\} \\
	&= q\, \{ \Lambda^{\frac{1}{2}} (\xi - \hat{\xi}) \}^\T \ee_{{H}}({H}^\T {H})\, \Lambda^{\frac{1}{2}} (\xi - \hat{\xi})
	= p\, \| \Lambda^{\frac{1}{2}} (\xi - \hat{\xi}) \|^2,
\end{aligned}
\]
where the last equality follows from $\ee({H}^\T {H}) = p I_q/q$.
Taking the expectation of \eqref{eq:dowson} with respect to ${H}$ and inserting the last display yields the required result.
\end{proof}

\begin{proof}[of Theorem~\ref{thm:CCLT_bound}]
Let $\pi_0$ denote the density function of $\law( {H} a \mid S^\T y, S^\T Z)  \ast \N_{\sigma^2}$ and let $\ee_0(\cdot)$ denote the expectation with respect to this distribution. Let $\upsilon\sim\pi_0$. By the definition of the Kullback-Leibler divergence,
\[
\begin{aligned}
	\Delta
	&= \ee_{0} \left\{
		\log \left(\frac{\pi_0(\upsilon)}{  \N[ \upsilon \mid {H} b_a, \,
	\{\gtr( \Lambda \hat{\Psi}) + \sigma^2 \} I_p ] }
	\right)\right\} \\
	&= \underbrace{\ee_{0} \left\{
		\log \left(\frac{\pi_0(\upsilon)}{  \N[ \upsilon \mid 0, \,
	\{\gtr( \Lambda \Psi) + \sigma^2 \} I_p ] }
	\right)\right\}}_{\Delta_1}
	+ \ee_{0} \left\{
		\log \left(\frac{  \N[ \upsilon \mid 0, \,
	\{\gtr( \Lambda \Psi) + \sigma^2 \} I_p ] }{  \N[ \upsilon \mid {H} b_a, \,
	\{\gtr( \Lambda \hat{\Psi}) + \sigma^2 \} I_p ] }
	\right)\right\}.
\end{aligned}
\]
Taking the expectation with respect to ${H}$ yields
\begin{equation} \label{eq:exp_Delta}
	\ee_{{H}}(\Delta) = \ee_{{H}}(\Delta_1)
	+ \ee_{{H}} \left[ \ee_{0} \left\{
		\log \left(\frac{  \N[ \upsilon \mid 0, \,
	\{\gtr( \Lambda \Psi) + \sigma^2 \} I_p ] }{  \N[ \upsilon \mid {H} b_a, \,
	\{\gtr( \Lambda \hat{\Psi}) + \sigma^2 \} I_p ] }
	\right)\right\} \right].
\end{equation}
Denote the expectation with respect to $\upsilon\sim \N[
	0, \,
	\{\gtr( \Lambda \Psi) + \sigma^2 \} I_p
]$ by $\ee_2(\cdot)$.
The mean and covariance of $\ee_H\{\law( {H} a \mid S^\T y, S^\T Z)  \ast \N_{\sigma^2}\}$
and $\N[0, \,
	\{\gtr( \Lambda \Psi) + \sigma^2 \} I_p ]$
are the same as confirmed in the proof of Lemma~\ref{lem:cclt},
and
the expectation of the logarithm of the Gaussian density only depends on the mean and covariance of $\upsilon$.
Therefore,
\begin{equation} \label{eq:Delta2_numerator}
	\ee_{{H}} \left[ \ee_{0} \left\{
		\log \left(\N[ \upsilon \mid 0, \,
	\{\gtr( \Lambda \Psi) + \sigma^2 \} I_p ]
	\right)\right\} \right]
	= \ee_{2} \left\{
		\log \left(\N[ \upsilon \mid 0, \,
	\{\gtr( \Lambda \Psi) + \sigma^2 \} I_p ]
	\right)\right\}.
\end{equation}
Also, expanding the square inside the Gaussian density and noting $\ee_0(\upsilon)=0$ yields
\begin{multline*}
	\ee_{{H}} \left[ \ee_{0} \left\{
		\log \left( \N[ \upsilon \mid {H} b_a, \,
	\{\gtr( \Lambda \hat{\Psi}) + \sigma^2 \} I_p ]
	\right)\right\} \right]
	\\ = \ee_{{H}} \left[ \ee_{0} \left\{
		\log \left( \N[ \upsilon \mid 0, \,
	\{\gtr( \Lambda \hat{\Psi}) + \sigma^2 \} I_p ]
	\right)
	+ \frac{\|{H} b_a\|^2}{ 2\{ \gtr( \Lambda \hat{\Psi}) + \sigma^2 \} }
	\right\} \right].
\end{multline*}
Again using that the logarithm of a Gaussian density only depends on the mean and covariance of $\nu$ provides
\begin{multline*}
	\ee_{{H}} \left[ \ee_{0} \left\{
		\log \left( \N[ \upsilon \mid {H} b_a, \,
	\{\gtr( \Lambda \hat{\Psi}) + \sigma^2 \} I_p ]
	\right)\right\} \right]
	\\
	\begin{aligned}
	&= \ee_{{H}} \left[ \ee_{2} \left\{
		\log \left( \N[ \upsilon \mid 0, \,
	\{\gtr( \Lambda \hat{\Psi}) + \sigma^2 \} I_p ]
	\right)
	+ \frac{\|{H} b_a\|^2}{ 2\{ \gtr( \Lambda \hat{\Psi}) + \sigma^2 \} }
	\right\} \right] \\
	&= \ee_{{H}} \left[ \ee_{2} \left\{
		\log \left( \N[ \upsilon \mid {H} b_a, \,
	\{\gtr( \Lambda \hat{\Psi}) + \sigma^2 \} I_p ]
	\right)
	\right\} \right],
	\end{aligned}
\end{multline*}
where the last equality follows from completing the square and $\ee_2(\upsilon)=0$.
Inserting the last display and \eqref{eq:Delta2_numerator} into \eqref{eq:exp_Delta}, and recalling the definition of the Kullback-Leibler divergence shows
\[
	\ee_{{H}}(\Delta) = \ee_{{H}}(\Delta_1) + \ee_{{H}}(\Delta_2).
\]

Both distributions in the Kullback-Leibler divergence $\Delta$
are equal to their respective distributions in the divergence in Theorem~\ref{thm:CCLT_bound} shifted by ${H} q^{1/2} \Lambda^{1/2} \xi = {M}^\T Z \xi$.
Since the Kullback-Leibler divergence is invariant to one-to-one transformations \citep[Corollary~4.1]{Kullback1951},
$\Delta$ equals the divergence in Theorem~\ref{thm:CCLT_bound}.
Also, $H$ is a deterministic function of ${M}^\T Z$ such that taking the expectation with respect to one or the other yields the same result.
Therefore, $\ee_{{H}}(\Delta)$ equals the left-hand side of Theorem~\ref{thm:CCLT_bound}.
The required result is thus $\ee_{{H}}(\Delta) \leq \delta_1 + \delta_2$ which
inserting Lemmas~\ref{lem:cclt} and \ref{lem:KLdiff} into the last display provides.
\end{proof}

\section{Proof of Theorem~\ref{thm:consistencyS}}
\label{sec:consistency_proof}

Let $P_{\gamma'} = X_{\gamma} (X_{\gamma}^\T X_{\gamma})^{-1} X_{\gamma}^\T$ denote the orthogonal projection onto the column space of $X_{\gamma}$.
The assumptions in Theorem~\ref{thm:consistencyS} in addition to \eqref{eq:eta_assum} are 
\begin{subequations} \label{eq:assum}
\begin{align}
\label{eq:prior_assum}
\pr(\gamma = \gamma^0) &> 0, \\
\label{eq:residual_assum}
\lim_{n\to\infty} \frac{\| (I_n - P_{\gamma}) X\beta^0 \|^2}{n} &> 0 \ \textnormal{for any $\gamma$ not containing $\gamma^0$}, \\
\label{eq:g_large_assum}
g_n &\to \infty, \\
\label{eq:g_small_assum}
\frac{\log g_n}{n} &\to 0.
\end{align}
\end{subequations}
Assumption \eqref{eq:prior_assum} is a basic prerequisite as otherwise $\pr(\gamma = \gamma^0\mid y) = 0$.
Assumption \eqref{eq:residual_assum} is analogous to Equation~A.4 from \citet{Fernandez2001}.
Previous literature \citep{Fernandez2001,Liang2008} required $g_n$ to grow appropriately with $n$, estimates $g_n$ via empirical Bayes, or places an appropriate prior on $g_n$
to obtain consistency.
We focus on the first case by assuming \eqref{eq:g_large_assum} and \eqref{eq:g_small_assum}.
Condition \eqref{eq:residual_assum} ensures that any model that does not contain the true one has posterior probability converging to zero. The fact that supersets of the true model are also discarded follows from  the $g$-prior, which favors smaller subsets.

\begin{lemma} \label{lem:orthTrick}
$P_{\gamma^0} - P_{\gamma} = {M} {M}^\T (P_{\gamma^0} - P_{\gamma})$.
\end{lemma}
\begin{proof}
Recall from \Sec{}\ref{sec:dataRot} that ${S}^\T X = 0_{(n-p)\times p}$ and $Q$ is orthogonal so that $Q Q^\T = I_n$. Therefore,
\[
	{M}{M}^\T X
	= {M}{M}^\T X + {S} \underbrace{\left( {S}^\T  X \right)}_{0_{(n-p)\times p}}
	= \left( {M}{M}^\T + {S}{S}^\T \right) X
	= \left( Q Q^\T \right) X
	= I_n X = X,
\]
where the third equality follows from $Q = \left({M},\ {S}\right)$.
Considering ${M}{M}^\T X = X$ columnwise and recalling $P_{\gamma} = X_\gamma (X_\gamma^\T X_\gamma)^{-1} X_\gamma^\T$
yields ${M} {M}^\T P_\gamma = P_\gamma$, for any $\gamma$ including $\gamma^0$.
\end{proof}

\begin{proof}[of Theorem~\ref{thm:consistencyS}]
Conditional on $\gamma$ and $\eta$, the set-up is a normal-normal model as both the prior ${\law(\beta_{\gamma}\mid\gamma)}$ from \eqref{eq:gprior} and the likelihood from \eqref{eq:lm} are Gaussian.
The corresponding marginal likelihood follows as
\begin{multline*}
	{\ddens}\left( y \mid \gamma, \eta \right)
	= \int {\ddens}\left( y \mid \beta_\gamma, \gamma, \eta \right)\, {\ddens}\left( \beta_\gamma \mid \gamma \right)\, d\beta_\gamma \\
	= \left( 2\pi\sigma^2 \right)^{-\frac{p}{2}}
	\left( g_n+1 \right)^{-\frac{|\gamma|}{2}}
	\exp\left\{
	-\frac{1}{2\sigma^2}\left(
		\|z\|^2
		- \frac{g_n}{g_n+1}
		z^\T P_{\gamma} z
	\right) \right\},
\end{multline*}
where
$z=y-\eta$
and $|\gamma|$ denotes the number of elements in $\gamma$.
The logarithm of the Bayes factor of the true model $\gamma^0$ over $\gamma$
conditional on $\eta$ is thus
\begin{equation} \label{eq:bf}
	\log \textsc{bf}_{\gamma^0:\gamma}
	= \log\left\{ \frac{ {\ddens}\left( y \mid \gamma^0, \eta \right)}{{\ddens}\left( y \mid \gamma, \eta \right)} \right\}
	=
	\frac{ |\gamma| - |\gamma^0| }{2}\, \log(g_n+1)
	+ \frac{g_n}{2\sigma^2(g_n+1)}\,
		h_\gamma(z),
\end{equation}
where $h_\gamma(z) = z^\T ( P_{\gamma^0} - P_{\gamma} ) z$.
By assumption \eqref{eq:prior_assum}, the required result follows if $\log \textsc{bf}_{\gamma^0:\gamma}\to\infty$ in probability, except for $\gamma=\gamma^0$ when $\log \textsc{bf}_{\gamma^0:\gamma^0}= 0$.

Since $z=y-\eta$, $z \sim \N(\nu,\, \sigma^2 I_n)$ where $\nu = X\beta^0 + \eta^0 - \eta$. Then, by Theorems~5.2a and 5.2c from \citet{Rencher2008}
and the fact that the trace of a projection matrix equals the dimensionality of its target space,
\begin{subequations} \label{eq:cond}
\begin{align}
\begin{split} \label{eq:cond_exp}
	\ee \left\{ h_\gamma(z) \mid \eta \right\}
	&= \sigma^2 \gtr ( P_{\gamma^0} - P_{\gamma} )
	+ \nu^\T ( P_{\gamma^0} - P_{\gamma} ) \nu \\
	&= \sigma^2 (|\gamma^0| - |\gamma|) + \nu^\T ( P_{\gamma^0} - P_{\gamma} ) \nu,
\end{split} \\
\begin{split} \label{eq:cond_var}
	\var \left\{ h_\gamma(z) \mid \eta \right\}
	&= 2\sigma^4 \gtr \left\{ ( P_{\gamma^0} - P_{\gamma} )^2
	\right\} + 4\sigma^2 \nu^\T ( P_{\gamma^0} - P_{\gamma} )^2 \nu \\
	&\leq 2\sigma^4 \left\{
		\gtr(P_{\gamma^0})
		+ \gtr(P_{\gamma})
	\right\}
	+ 4\sigma^2 \| ( P_{\gamma^0} - P_{\gamma} ) \nu\|^2 \\
	&= 2\sigma^4 (|\gamma^0| + |\gamma|)
	+ 4\sigma^2 \| ( P_{\gamma^0} - P_{\gamma} ) \nu\|^2.
\end{split}
\end{align}
\end{subequations}
We analyze the asymptotic behavior of $h_\gamma(z)$ by bounding this expectation and variance.

The first term of each right-hand side in \eqref{eq:cond} is independent of $n$.
Let us bound the second terms.
Inserting
$\nu =X \beta^0 + \zeta$ where $\zeta = \eta^0 -  \eta$
and expanding the square yields
\[
	\nu^\T ( P_{\gamma^0} - P_{\gamma} ) \nu
	= (X\beta^0)^\T (P_{\gamma^0} - P_{\gamma}) X\beta^0 +  2 \zeta^\T  (P_{\gamma^0} - P_{\gamma}) X\beta^0 + \zeta^\T (P_{\gamma^0} - P_{\gamma}) \zeta.
\]
Inserting $P_{\gamma^0} X \beta^0 = X\beta^0$ and Lemma~\ref{lem:orthTrick} provides
\[
\begin{aligned}
	\nu^\T ( P_{\gamma^0} - P_{\gamma} ) \nu
	&=
  (X\beta^0)^\T (I_n - P_{\gamma}) X\beta^0 +  2 \zeta^\T {M} {M}^\T (P_{\gamma^0} - P_{\gamma}) X\beta^0 + \zeta^\T (P_{\gamma^0} - P_{\gamma}) \zeta \\
&=
\| (I_n - P_{\gamma}) X\beta^0\|^2 +  2 \zeta^\T {M}{M}^\T (I_n - P_{\gamma}) X\beta^0 + \zeta^\T {M}{M}^\T (P_{\gamma^0} - P_{\gamma}) \zeta.
\end{aligned}
\]
Applying the Cauchy-Schwarz inequality and $|\zeta^\T {M}{M}^\T (P_{\gamma^0} - P_{\gamma}) \zeta| \leq \zeta^\T {M}{M}^\T \zeta = \|{M}^\T \zeta\|^2$,
\begin{align*}
	\nu^\T ( P_{\gamma^0} - P_{\gamma} ) \nu &\geq
	\| (I_n - P_{\gamma}) X\beta^0\|^2 -   2 \| {M}^\T \zeta \|\, \|{M}^\T (I_n - P_{\gamma}) X\beta^0\|   -  \|{M}^\T \zeta\|^2, \\
	\nu^\T ( P_{\gamma^0} - P_{\gamma} ) \nu &\leq
	\| (I_n - P_{\gamma}) X\beta^0\|^2 +   2 \| {M}^\T \zeta \|\, \|{M}^\T (I_n - P_{\gamma}) X\beta^0\|   +  \|{M}^\T \zeta\|^2.
\end{align*}
Since the columns of $M$ form an orthonormal basis for the column space of $X$,
${\|{M}^\T (I_n - P_{\gamma}) X\beta^0\|} = {\|(I_n - P_{\gamma}) X\beta^0\|}$ such that
\begin{subequations}
\begin{align}
\begin{split} \label{eq:exp_bound_lower}
	\nu^\T ( P_{\gamma^0} - P_{\gamma} ) \nu &\geq
	\| (I_n - P_{\gamma}) X\beta^0\|^2 -   2 \| {M}^\T \zeta \|\, \|(I_n - P_{\gamma}) X\beta^0\|   -  \|{M}^\T \zeta\|^2 \\
	&=
\left\{ \| (I_n - P_{\gamma}) X\beta^0\| -   2 \| {M}^\T \zeta \|\right\}\, \| (I_n - P_{\gamma}) X\beta^0\|  -  \| {M}^\T \zeta \|^2,
\end{split} \\
\begin{split} \label{eq:exp_bound_upper}
	\nu^\T ( P_{\gamma^0} - P_{\gamma} ) \nu &\leq
	\| (I_n - P_{\gamma}) X\beta^0\|^2 +   2 \| {M}^\T \zeta \|\, \|(I_n - P_{\gamma}) X\beta^0\|   +  \|{M}^\T \zeta\|^2 \\
	&=
\left\{ \| (I_n - P_{\gamma}) X\beta^0\| +   2 \| {M}^\T \zeta \|\right\}\, \| (I_n - P_{\gamma}) X\beta^0\|  +  \| {M}^\T \zeta \|^2.
\end{split}
\end{align}
\end{subequations}

For the second term of the right-hand side in \eqref{eq:cond_var},
consider $\nu =X \beta^0 + \zeta$ and
\[
	\| ( P_{\gamma^0} - P_{\gamma} ) \nu\|
	= \| (I_n - P_{\gamma}) X \beta^0 + (P_{\gamma^0} - P_{\gamma}) \zeta \|.
\]
By the triangle inequality,
\[
\begin{aligned}
	\| ( P_{\gamma^0} - P_{\gamma} ) \nu\|
	&\leq \| (P_{\gamma^0} - P_{\gamma}) X \beta^0 \| + \| (P_{\gamma^0} - P_{\gamma}) \zeta \| \\
	&= \| (I_n - P_{\gamma}) X \beta^0 \| + \| (P_{\gamma^0} - P_{\gamma}) {M}{M}^\T \zeta \|,
\end{aligned}
\]
where the equality follows from $P_{\gamma^0} X \beta^0 = X\beta^0$
and Lemma~\ref{lem:orthTrick}.
Also,
$\| (P_{\gamma^0} - P_{\gamma}) {M}{M}^\T \zeta \| \leq \| {M}{M}^\T \zeta \| = \| {M}^\T \zeta \|$ since ${M}^\T {M} = I_p$.
Therefore,
\[
	\| ( P_{\gamma^0} - P_{\gamma} ) \nu\|
	\leq \| (I_n - P_{\gamma}) X \beta^0 \| + \| {M}^\T \zeta \|.
\]
Inserting into \eqref{eq:cond_var} provides
\begin{equation} \label{eq:var_bound}
\begin{aligned}
\var \left\{ h_\gamma(z )\mid \eta \right\}^{\frac{1}{2}}
&\leq \left\{ 2\sigma^4 (|\gamma^0| + |\gamma|) +  4 \sigma^2 \| ( P_{\gamma^0} - P_{\gamma} ) \nu\|^2 \right\}^{\frac{1}{2}} \\
& \le 2^{\frac{1}{2}}\,  \sigma^2 (|\gamma^0| + |\gamma|)^{\frac{1}{2}}
+ 2 \sigma \| ( P_{\gamma^0} - P_{\gamma} ) \nu\| \\
&\leq 2^{\frac{1}{2}} \,  \sigma^2 (|\gamma^0| + |\gamma|)^{\frac{1}{2}} + 2\sigma\left\{ \| (I_n - P_{\gamma}) X \beta^0\|  + \| {M}^\T \zeta \| \right\}. 
\end{aligned}
\end{equation}

Since $\zeta = \eta^0 - \eta$, assumptions \eqref{eq:eta_assum} and \eqref{eq:g_small_assum} imply
\begin{equation} \label{eq:zeta_assum}
	\frac{\| M^\T \zeta \|^2}{\log g_n} \to 0,
	\qquad
	\frac{\| M^\T \zeta \|^2}{n} \to 0;
\end{equation}
in probability.
Let us consider $\gamma\ne\gamma^0$ that contain $\gamma^0$, that is $\gamma^0 \subsetneq \gamma$, and $\gamma$ that do not contain $\gamma^0$, that is $\gamma^0 \not\subset \gamma$, separately.

First, consider the case where $\gamma$ does not contain $\gamma^0$.
Assumption \eqref{eq:residual_assum}, \eqref{eq:cond_exp}, \eqref{eq:exp_bound_lower}, and \eqref{eq:zeta_assum} imply
$\ee \left\{ h_\gamma(z) \mid \eta \right\}/\| (I_n - P_{\gamma}) X \beta^0 \| \to \infty$.
On the other hand,
$\lim_{n\to\infty} {\var \{ h_\gamma(z )\mid \eta \}^{1/2}} /\| (I_n - P_{\gamma}) X \beta^0 \| \leq 2\sigma$ by \eqref{eq:residual_assum}, \eqref{eq:var_bound}, and \eqref{eq:zeta_assum}.
Therefore,
$\lim_{n\to\infty} h_\gamma(z\mid \eta )/n > 0$
with probability tending to one
by Chebyshev's inequality and \eqref{eq:residual_assum}.
Under assumption \eqref{eq:g_small_assum}, it then follows from \eqref{eq:bf} that $\log \textsc{bf}_{\gamma^0:\gamma}\to\infty$ in probability. 

Next, consider the case where $\gamma$ contains $\gamma_0$. In this setting, $P_{\gamma} X\beta^0 = X\beta^0$ and thus ${(I_n - P_{\gamma})} X\beta^0  = 0_{n\times 1}$.
Therefore,
\eqref{eq:cond_exp} with \eqref{eq:exp_bound_upper}, and \eqref{eq:var_bound} reduce to
\begin{align*}
	\ee \left\{ h_\gamma(z) \mid \eta \right\}
	&\leq \sigma^2 (|\gamma^0| - |\gamma|)+\| {M}^\T \zeta \|^2, \\
	\var \left\{ h_\gamma(z )\mid \eta \right\}^{\frac{1}{2}}
	&\leq 2^{\frac{1}{2}}\,  \sigma^2 (|\gamma^0| + |\gamma|)^{\frac{1}{2}} + 2\sigma\,\| {M}^\T \zeta \|.
\end{align*}
Chebyshev's inequality and \eqref{eq:zeta_assum}
provide thus
$\lim_{n\to\infty} {h_\gamma(z\mid \eta )} / \log g_n = 0$
with probability tending to one.  We conclude from \eqref{eq:bf} that  $\textsc{bf}_{\gamma^0:\gamma}\to\infty$ in probability because of assumption \eqref{eq:g_large_assum} and $|\gamma| > | \gamma_0|$.

We have shown $\textsc{bf}_{\gamma^0:\gamma}\to\infty$ whenever $\gamma\ne\gamma^0$. The required result follows from this result as noted earlier in this proof.
\end{proof}

\section{Vector approximate message passing} \label{sec:vamp}

\subsection{Derivation}

\begin{figure}[htb]
  \centering
  \psfragfig{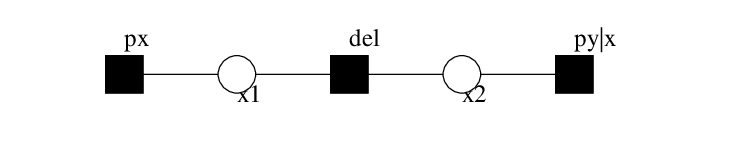}{
  \newcommand{\sz}{0.8}
  \psfrag{px}[b][Bl][\sz]{$\pi(\beta)$}
 \psfrag{x1}[t][Bl][\sz]{$\beta$}
 \psfrag{del}[b][Bl][\sz]{$\delta(\beta-\tilde{\beta})$}
 \psfrag{x2}[t][Bl][\sz]{$\tilde{\beta}$}
 \psfrag{py|x}[b][Bl][\sz]{$\N(y \mid X \tilde{\beta},\ \sigma^2 I_n)$}
 }
  \caption{The factor graph representation of \eqref{eq:factor_graph}.
           The squares and circles are factor and variable nodes, respectively.
           This figure is an edited version of Fig.~1 from \citet{Rangan2016}.}
  \label{fig:factor_graph}
\end{figure}

To give a motivation for the steps of vector approximate message passing in Algorithm~\ref{alg:vamp} on page~\pageref{alg:vamp},
we derive the algorithm as an approximation to sum-product message passing \citep[\Sec{}8.4.4]{Bishop2006} similar to what is done in \citet[\Sec{}III-B]{Rangan2016}.
Consider the linear model
$y\sim\N(X\beta, \sigma^2 I_n)$
where $y$ is an $n$-dimensional vector of observations, $X$ an $n\times p$ design matrix, $\beta$ a $p$-dimensional vector of parameters, and $\sigma^2$ the error variance.
We assume that the entries of $\beta$ are a priori independent such that $\pi(\beta) = \prod_{j=1}^p \pi(\beta_j)$.
The goal is to approximate the posterior
\begin{equation} \label{eq:factor_graph}
\begin{split}
	\pi(\beta\mid y) \propto
	\pi(\beta)\, \pi(y\mid \beta)
	&= \pi(\beta)\, \N(y \mid X\beta,\ \sigma^2 I_n) \\
	&= \pi(\beta)\, \delta(\beta-\tilde{\beta})\, \N(y \mid X \tilde{\beta},\ \sigma^2 I_n),
\end{split}
\end{equation}
where $\delta$ is the Dirac delta function and $\tilde{\beta}$ is thus a copy of $\beta$.
This copying of $\beta$ gives rise to an extra variable node in the corresponding factor graph in Fig.~\ref{fig:factor_graph}.

Let $\mu_{\pi\to\beta}$ and $\mu_{\delta\to\beta}$ denote the messages to the variable node $\beta$, $\mu_{\delta\to\tilde{\beta}}$ and $\mu_{\N\to\tilde{\beta}}$ the messages to the variable node $\tilde{\beta}$,
and $\mu_{\beta\to\delta}$ and $\mu_{\tilde{\beta}\to\delta}$ the messages to the factor node $\delta(\beta-\tilde{\beta})$.
By the general expression for a message from a factor to a variable node \citep[Equation~8.69]{Bishop2006},
\begin{equation} \label{eq:copy_message}
\begin{split}
\mu_{\delta\to\tilde{\beta}}(\tilde{\beta}) &= \int \delta(\beta - \tilde{\beta}) \mu_{\beta\to \delta}(\beta) d\beta = \mu_{\beta\to\delta}(\tilde{\beta}), \\
\mu_{\delta\to\beta}(\beta) &= \int \delta(\beta - \tilde{\beta}) \mu_{\tilde{\beta}\to \delta}(\tilde{\beta}) d\tilde{\beta} = \mu_{\tilde{\beta}\to\delta}(\beta).
\end{split}
\end{equation}
The beliefs at the variable nodes are the products of the incoming messages,
\[
\begin{split}
b(\beta) &\propto \mu_{\pi\to\beta}(\beta)\, \mu_{\delta\to\beta}(\beta)
= \pi(\beta)\, \mu_{\delta\to\beta}(\beta),\\
b(\tilde{\beta}) &\propto \mu_{\delta\to\tilde{\beta}}(\tilde{\beta})\, \mu_{\N \to \tilde{\beta}}(\tilde{\beta})
= \mu_{\beta\to\delta}(\tilde{\beta})\, \N(y \mid X \tilde{\beta},\ \sigma^2 I_n);
\end{split}
\]
where the last equality uses \eqref{eq:copy_message}.
Combining these beliefs with the general expression for a message from a variable to a factor node \citep[Equation~8.66]{Bishop2006} and Fig.~\ref{fig:factor_graph} yields
\[
\begin{split}
	\mu_{\beta\to \delta}(\beta) &=
	\mu_{\pi\to\beta}(\beta) \propto \frac{b(\beta)}{\mu_{\delta\to\beta}(\beta)}, \\
	\mu_{\tilde{\beta} \to \delta}(\tilde{\beta}) &= \mu_{\N\to\tilde{\beta}}(\tilde{\beta})
	\propto \frac{b(\tilde{\beta})}{\mu_{\delta\to\tilde{\beta}}(\tilde{\beta})}
	= \frac{b(\tilde{\beta})}{\mu_{\beta\to\delta}(\tilde{\beta})};
\end{split}
\]
where the last equality follows from \eqref{eq:copy_message}.

The last two displays provide a message-passing algorithm. Initialize $\mu_{\delta\to\beta}(\beta)$. Then, iterate the updates
\begin{subequations} \label{eq:bp_alg}
\begin{align}
	\label{eq:bp_alg_a}
	b(\beta) &\propto \pi(\beta)\, \mu_{\delta\to\beta}(\beta), \\
	\label{eq:bp_alg_b}
	\mu_{\beta\to \delta}(\beta) &\propto \frac{b(\beta)}{\mu_{\delta\to\beta}(\beta)}, \\
	\label{eq:bp_alg_c}
	b(\tilde{\beta}) &\propto\mu_{\beta\to\delta}(\tilde{\beta})\, \N(y \mid X \tilde{\beta},\ \sigma^2 I_n), \\
	\label{eq:bp_alg_d}
	\mu_{\delta\to\beta}(\tilde{\beta}) = \mu_{\tilde{\beta} \to \delta}(\tilde{\beta}) &\propto \frac{b(\tilde{\beta})}{\mu_{\beta\to\delta}(\tilde{\beta})},
\end{align}
\end{subequations}
where the last equality is from \eqref{eq:copy_message}.
Since the graph in Fig.~\ref{fig:factor_graph} is a tree, the beliefs $b(p)$ converge to the exact posterior $\pi(\beta\mid y)$ after one iteration.
This exact algorithm can however be expensive to compute for certain $\pi(\beta)$ if $p$ is large.
Vector approximate message passing approximates \eqref{eq:bp_alg} to reduce computational cost:

Initialize $\mu_{\delta\to\beta}(\beta) = \N(\beta\mid r_0,\ t_0^2 I_p)$.
At the $k$th iteration,
approximate $b(\beta)$ by ${\N(\beta\mid \hat{\beta}_k,\ s_k^2 I_p)}$
where $\hat{\beta}_k = \ee_{b(\beta)}(\beta)$ and $s_k^2 = \mathrm{Tr}\{ \cov_{b(\beta)}(\beta) \} / p$.
Applying \eqref{eq:bp_alg_a} provides Step~\ref{alg:vamp_a} of Algorithm~\ref{alg:vamp}.

Since $\mu_{\delta\to\beta}(\beta) \approx \N(\beta\mid r_k,\ t_k^2 I_p)$ and $b(\beta) \approx \N(\beta\mid \hat{\beta}_k,\ s_k^2 I_p)$, the resulting approximation to $\mu_{\beta\to \delta}(\beta)$ is Gaussian too by \eqref{eq:bp_alg_b}.
Denote this Gaussian approximation by $\N(\tilde{\beta}\mid \tilde{r}_k,\ \tilde{t}_k^2 I_p)$. Step~\ref{alg:vamp_b} states the update equations for $\tilde{r}_k$ and $\tilde{t}_k^2$ derived from \eqref{eq:bp_alg_b}.

With $\mu_{\beta\to \delta}(\tilde{\beta})\approx \N(\tilde{\beta}\mid \tilde{r}_k,\ \tilde{t}_k^2 I_p)$,
$b(\tilde{\beta})$ from \eqref{eq:bp_alg_c} is Gaussian too.
We further approximate $b(\tilde{\beta})$ by requiring its covariance to be proportional to the identity matrix.
Let $b(\tilde{\beta})\approx {\N(\tilde{\beta}\mid \hat{\tilde{\beta}}_k,\ \tilde{s}_k^2 I_p)}$.
The updates follow from \eqref{eq:bp_alg_c} as
\begin{subequations}
\begin{align}
	\label{eq:beta_expensive}
	\hat{\tilde{\beta}}_k &= \left(
		\tilde{t}_k^2 X^\T X + \sigma^2 I_p
	\right)^{-1} \left(
		\tilde{t}_k^2 X^\T y + \sigma^2 \tilde{r}_k
	\right), \\
	\label{eq:s_expensive}
	\tilde{s}^2_k &= \frac{\sigma^2 \tilde{t}^2_k}{p} \,\mathrm{Tr}\left\{\left(
		\tilde{t}_k^2 X^\T X + \sigma^2 I_p
	\right)^{-1} \right\}.
\end{align}
\end{subequations}
These involve an inversion of a $p\times p$ matrix which is expensive to compute if $p$ is large. We can however rewrite these expressions to make their computation faster.

Let $X = UDV^\T$ denote a singular-value decomposition with an $n\times\min(n,p)$ matrix $U$, a $\min(n,p)\times\min(n,p)$ diagonal matrix $D$, and $V$ an $p\times\min(n,p)$ matrix such that $U^\T U = V^\T V = I_{\min(n,p)}$.
Substituting $X = UDV^\T$ yields
\begin{equation} \label{eq:woodbury}
\begin{aligned}
	\left(
		\tilde{t}_k^2 X^\T X + \sigma^2 I_p
	\right)^{-1}
	&= \left(
		\tilde{t}_k^2 V D^2 V^\T + \sigma^2 I_p
	\right)^{-1}
	= \frac{1}{\sigma^2} \left(
		\frac{\tilde{t}_k^2}{\sigma^2} V D^2 V^\T + I_p
	\right)^{-1} \\
	&= \frac{1}{\sigma^2} \left\{
		I_p - V\left(
		\frac{\sigma^2}{\tilde{t}_k^2} D^{-2} + V^\T V
		\right)^{-1} V^\T
	\right\} \\
	&= \frac{1}{\sigma^2} \left[
		I_p - V\left\{
		\frac{\sigma^2}{\tilde{t}_k^2} D^{-2} + I_{\min(n,p)}
		\right\}^{-1} V^\T
	\right],
\end{aligned}
\end{equation}
where $D^2 = DD$, $D^{-2}=D^{-1}D^{-1}$ and the third equality follows from the Woodbury matrix identity.
Substituting $X^\T = VDU^\T$ and \eqref{eq:woodbury} provide
\[
\begin{aligned}
	\left(
		\tilde{t}_k^2 X^\T X + \sigma^2 I_p
	\right)^{-1} \tilde{t}_k^2 X^\T y
	&= \frac{\tilde{t}_k^2}{\sigma^2} \left[
		I_p - V\left\{
		\frac{\sigma^2}{\tilde{t}_k^2} D^{-2} + I_{\min(n,p)}
		\right\}^{-1} V^\T
	\right] VDU^\T y \\
	&= \frac{\tilde{t}_k^2}{\sigma^2} \left[
		V - V\left\{
		\frac{\sigma^2}{\tilde{t}_k^2} D^{-2} + I_{\min(n,p)}
		\right\}^{-1} V^\T V
	\right] DU^\T y \\
	&= \frac{\tilde{t}_k^2}{\sigma^2} V\left[
		I_{\min(n,p)} - \left\{
		\frac{\sigma^2}{\tilde{t}_k^2} D^{-2} + I_{\min(n,p)}
		\right\}^{-1}
	\right] DU^\T y,
\end{aligned}
\]
where the last equality uses $V^\T V = I_{\min(n,p)}$.
Since the expression inside the square brackets consists only of diagonal matrices, we can write it as a single fraction to obtain
\[
\begin{aligned}
\left(
		\tilde{t}_k^2 X^\T X + \sigma^2 I_p
	\right)^{-1} \tilde{t}_k^2 X^\T y
	&= \frac{\tilde{t}_k^2}{\sigma^2} V\left[
		\frac{\sigma^2}{\tilde{t}_k^2} D^{-2} \left\{
		\frac{\sigma^2}{\tilde{t}_k^2} D^{-2} + I_{\min(n,p)}
		\right\}^{-1}
	\right] DU^\T y \\
	&= \frac{\tilde{t}_k^2}{\sigma^2} V
		\left\{ I_{\min(n,p)} + \frac{\tilde{t}_k^2}{\sigma^2} D^{2} \right\}^{-1}
	 DU^\T y \\
	&= V
		\left\{ \frac{\sigma^2}{\tilde{t}_k^2} I_{\min(n,p)} + D^2 \right\}^{-1}
	DU^\T y,
\end{aligned}
\]
where the second equality follows from multiplying both the numerator and the denominator of the diagonal-matrices faction by $(\tilde{t}_k^2 / \sigma^2) D^{2}$.
Combining the last display with \eqref{eq:beta_expensive} and \eqref{eq:woodbury} provides
\begin{equation} \label{eq:beta_cheap}
\begin{aligned}
	\hat{\tilde{\beta}}_k &= V
		\left\{ \frac{\sigma^2}{\tilde{t}_k^2} I_{\min(n,p)} + D^2 \right\}^{-1}
	DU^\T y + \left[
		I_p - V\left\{
		\frac{\sigma^2}{\tilde{t}_k^2} D^{-2} + I_{\min(n,p)}
		\right\}^{-1} V^\T \right] \tilde{r}_k \\
	&=
	\tilde{r}_k +
	V
		\left\{ \frac{\sigma^2}{\tilde{t}_k^2} I_{\min(n,p)} + D^2 \right\}^{-1}
	DU^\T y - 
		V\left\{
		\frac{\sigma^2}{\tilde{t}_k^2} D^{-2} + I_{\min(n,p)}
		\right\}^{-1} V^\T \tilde{r}_k \\
		&=
	\tilde{r}_k +
	V
		\left\{ \frac{\sigma^2}{\tilde{t}_k^2} I_{\min(n,p)} + D^2 \right\}^{-1}
	DU^\T y -
		 V
		\left\{ \frac{\sigma^2}{\tilde{t}_k^2} I_{\min(n,p)} + D^2 \right\}^{-1} D^2 V^\T \tilde{r}_k \\
	&=
	\tilde{r}_k +
	V
		\left\{ \frac{\sigma^2}{\tilde{t}_k^2} I_{\min(n,p)} + D^2 \right\}^{-1} \left(
	DU^\T y -
		D^2 V^\T \tilde{r}_k \right) \\
		&=
	\tilde{r}_k +
	V
		\left( \frac{\sigma^2}{\tilde{t}_k^2} D^{-1} + D \right)^{-1} \left(
	U^\T y -
		D V^\T \tilde{r}_k \right),
\end{aligned}
\end{equation}
where we used that $D$ is a diagonal matrix.
This update for $\hat{\tilde{\beta}}_k$ only involves matrix multiplications and inversions of diagonal matrices.

For $\tilde{s}^2_k$, substitute \eqref{eq:woodbury} into \eqref{eq:s_expensive} such that
\[
\begin{aligned}
	\tilde{s}^2_k &= \frac{\tilde{t}^2_k}{p} \mathrm{Tr} \left[
		I_p - V\left\{
		\frac{\sigma^2}{\tilde{t}_k^2} D^{-2} + I_{\min(n,p)}
		\right\}^{-1} V^\T
	\right] \\
	&= \tilde{t}^2_k \left(
		1 - \frac{1}{p} \mathrm{Tr} \left[
			\left\{
		\frac{\sigma^2}{\tilde{t}_k^2} D^{-2} + I_{\min(n,p)}
		\right\}^{-1} V^\T V
		\right]
	\right),
\end{aligned}
\]
where the last equality uses that the trace is invariant under cyclic permutations.
Since $V^\T V = I_{\min(n,p)}$, the last expression reduces to
\[
\begin{aligned}
	\tilde{s}^2_k
	&= \tilde{t}^2_k \left(
		1 - \frac{1}{p} \mathrm{Tr} \left[
			\left\{
		\frac{\sigma^2}{\tilde{t}_k^2} D^{-2} + I_{\min(n,p)}
		\right\}^{-1}
		\right]
	\right) \\
	&= \tilde{t}^2_k \left[
		1 - \frac{1}{p} \, \mathrm{Tr} 
			\left\{ D\,
		\left(\frac{\sigma^2}{\tilde{t}_k^2} D^{-1} + D \right)^{-1}
		\right\}
	\right], \\
\end{aligned}
\]
where the last equality uses that the argument of the trace is diagonal.
This display with \eqref{eq:beta_cheap} constitutes Step~\ref{alg:vamp_c}.

Recall $\mu_{\beta\to \delta}(\tilde{\beta})\approx \N(\tilde{\beta}\mid \tilde{r}_k,\ \tilde{t}_k^2 I_p)$ and $b(\tilde{\beta})\approx \N(\tilde{\beta}\mid \hat{\tilde{\beta}}_k,\ \tilde{s}_k^2 I_p)$.
We would like to update $\mu_{\delta\to\beta}(\beta) \approx \N(\beta\mid r_{k+1},\ t_{k+1}^2 I_p)$ where we have incremented the iteration counter.
Step~\ref{alg:vamp_d} follows now from \eqref{eq:bp_alg_d} in the same way as Step~\ref{alg:vamp_b} followed from \eqref{eq:bp_alg_b}.

\vspace*{12pt}
\begin{minipage}{\textwidth}
\begin{algo} \label{alg:vamp}
Vector approximate message passing. \\
Input: Data $(y,\ X)$
\begin{enumerate}[1.]
		\item \label{alg:svd}
		Compute the singular-value decomposition $X=U D V^\T$.
		\item
		Initialize $r_0$ and $t^2_0$.
		\item
		For $k=0,\dots,K$ do:
		\begin{enumerate}
			\item \label{alg:vamp_a}
			Set
			$\hat{\beta}_{k,j} = \ee(\beta_j\mid r_{k,j},\, t_k^2)$ and $s^2_k = \sum_{j=1}^p \var(\beta_j\mid r_{k,j},\, t_k^2) / p$
			where the density of $\beta_j$ is proportional to $\pi(\beta_j) \, \N(\beta_j\mid r_{k,j},\, t_k^2)$
			for $j=1,\dots,p$.
			\item \label{alg:vamp_b}
			Set $1/\tilde{t}^2_k = 1/s^2_k - 1/t^2_k$ and $\tilde{r}_k = (t^2_k\hat{\beta}_k - s^2_k r_k)/(t^2_k - s^2_k)$.
			\item \label{alg:vamp_c}
			Set $\hat{\tilde{\beta}}_k = \tilde{r}_k +
	V
		( \sigma^2 D^{-1} / \tilde{t}_k^2 + D )^{-1} (
	U^\T y -
		D V^\T \tilde{r}_k )$
		and \\
		$\tilde{s}^2_k = \tilde{t}^2_k\, [
		1 - \mathrm{Tr} 
			\{ D\,
		(\sigma^2 D^{-1} / \tilde{t}_k^2 + D )^{-1}
		\} / p
	]$.
	\item \label{alg:vamp_d}
			Set $1/t^2_{k+1} = 1/\tilde{s}^2_k - 1/\tilde{t}^2_k$ and $r_{k+1} = (\tilde{t}^2_k\hat{\tilde{\beta}}_k - \tilde{s}^2_k \tilde{r}_k)/(\tilde{t}^2_k - \tilde{s}^2_k)$.
		\end{enumerate}
	\end{enumerate}
Output: Approximate posterior $\N(\hat{\beta}_K,\ s_K^2 I_p)$
\end{algo}
\end{minipage}

\subsection{Computational complexity} \label{sec:vamp_comp}

The computational complexity of the singular-value decomposition is $O\{n\, p\, \min(n,p)\}$ \citep[\Sec{}I-E]{Rangan2016}.
The steps inside each iteration are $O(p)$ except for Step~\ref{alg:vamp_c} which is $O\{p\,\min(n,p)\}$ if $U^\T y$ is precomputed.
The computational complexity of Algorithm~\ref{alg:vamp} is thus $O\{(n+K)\, p\, \min(n,p)\}$.

In practice, we do not always run Algorithm~\ref{alg:vamp} for all $K$ iterations. We stop it once the innovation $\|\hat{\beta}_k - \hat{\beta}_{k-1}\|^2$ becomes small enough, indicating convergence.

\subsection{Estimating $\sigma^2$}
\label{sec:vamp_sigma}

So far, we have treated $\sigma^2$ as fixed and known.
As \Sec{}\ref{sec:sigma} notes, applications like those in \Sec{}\ref{sec:diabetes} and \Sec{}\ref{sec:snp} often require estimation of $\sigma^2$
and methods available for Step~2 of Algorithm~\ref{alg:irga} often provide such estimation.
For instance,
\citet{Vila2011} detail how $\sigma^2$ can be estimated when using approximate message passing.
We add a step to Algorithm~\ref{alg:vamp} to estimate $\sigma^2$ when required:
Consider the prior $1/\sigma^2 \sim \Ga(a_0,b_0)$ for some shape parameter $a_0$ and rate parameter $b_0$. Then, the full conditional posterior for $1/\sigma^2$ of Algorithm~\ref{alg:vamp} at iteration $k$ is
\[
	\frac{1}{\sigma^2} \mid \hat{\beta}_k \sim \Ga\left(
		a_0 + \frac{n}{2},\,
		b_0 + \frac{\| y - X \hat{\beta}_k \|^2}{2}
	\right).
\]
At each iteration, we update $\sigma^2$ such that $1/\sigma^2$ matches the mean of this full conditional:
\[
	\sigma^2_k = \frac{b_0 + \| y - X \hat{\beta}_k \|^2/2}{a_0 + n/2}
	\quad (k=1,\dots,K),
\]
between Steps~\ref{alg:vamp_a} and \ref{alg:vamp_b} of Algorithm~\ref{alg:vamp}.

\subsection{Dampened updates} \label{sec:dampening}

If vector approximate message passing fails to converge, which can happen for certain matrices $X$ which have a challenging collinearity structure, damping of updates can induce convergence, like it does in approximate message passing \citep{Rangan2014}. In this article, we only dampen the updates for the SNP application in \Sec{}\ref{sec:snp} to ensure convergence.

Let $\rho\in (0, 1]$ denote the damping constant with $\rho = 1$ representing no damping. Then,
the dampened version of Algorithm~\ref{alg:vamp} follows by replacing
Steps~\ref{alg:vamp_a} and \ref{alg:vamp_c} by
\[
\begin{aligned}
\hat{\beta}_{k,j} &= (1-\rho)\, \hat{\beta}_{k-1,j} + \rho\, \ee(\beta_j\mid r_{k,j},\, t_k^2), \\
s^2_k &= (1-\rho)\, s^2_{k-1} + \rho\, \sum_{j=1}^p \var(\beta_j\mid r_{k,j},\, t_k^2) / p;
\end{aligned}
\]
and
\[
\begin{aligned}
\hat{\tilde{\beta}}_k &= (1-\rho)\,\hat{\tilde{\beta}}_{k-1} +
\rho\,
\{ \tilde{r}_k
+
	V
		( \sigma^2 D^{-1} / \tilde{t}_k^2 + D )^{-1} (
	U^\T y -
		D V^\T \tilde{r}_k ) \}, \\
		\tilde{s}^2_k &= (1-\rho)\,\tilde{s}^2_{k-1} + \rho\, \tilde{t}^2_k\, [
		1 - \mathrm{Tr} 
			\{ D\,
		(\sigma^2 D^{-1} / \tilde{t}_k^2 + D )^{-1}
		\} / p
	];
\end{aligned}
\]
respectively, for $k>0$.

\section{Debiased lasso}
\label{sec:lasso}

Consider the linear model
$y\sim\N(X\beta, \sigma^2 I_n)$ as in \Sec{}\ref{sec:vamp}
with the spike-and-slab prior
$\beta_j\sim \lambda\, \N(0,\,\psi) + (1-\lambda)\,\delta(0)$ independently for $j=1,\dots,p$.
Denote the lasso estimator of $\beta$ by $\hat{\beta}^\text{lasso}(y)$: Here, we use the smallest lasso regularization parameter that results in at most $\lfloor \lambda p \rfloor$ nonzero coefficients
where $\lambda p$ is the number of expected nonzero elements in $\beta$ under its spike-and-slab prior.
The lasso algorithm from \citet{Efron2004} allows for efficient computation of $\hat{\beta}^\text{lasso}(y)$ under this constraint on the regularization parameter.

As the number of predictors is less than the sample size in \Sec{}\ref{sec:diabetes},
we assume $p\leq n$.
Then, we can set the matrix $M$ in \citet{Javanmard2013}
equal to $\hat{\Sigma}^{-1}$ where $\hat{\Sigma} = X^\T X / n$. The debiased lasso estimator follows as
\citep[Equation~5]{Javanmard2013}
\[
	\hat{\beta}^\text{unbiased}(y) = \hat{\beta}^\text{lasso}(y) + \frac{1}{n} M X^\T \left\{ y - \hat{\beta}^\text{lasso}(y) \right\}.
\]
Theorem~2.1 from \citet{Javanmard2013} implies
\[
	\beta \mid y \sim \mathcal{N}\left\{
		\hat{\beta}^\text{unbiased}(y),\,
		\frac{\sigma^2}{n}\, M \hat{\Sigma} M^\T
	\right\},
\]
as posterior approximation based on the debiased lasso.

The error variance $\sigma^2$ is unknown in the application from \Sec{}\ref{sec:diabetes}. We therefore estimate it by
\[
	\frac{b_0 + \| y - X \hat{\beta}^\text{unbiased}(y) \|^2/2}{a_0 + n/2},
\]
analogously to \Sec{}\ref{sec:vamp_sigma}.

\section{Laplace approximation for \Sec{}\ref{sec:nonparam_covariates}}
\label{sec:gauss_newton}

We follow \citet[\Sec{}2.3]{Steinberg2014}.
Recall $F = \{f(z_1),\dots,f(z_n)\}^\T$.
Since $f$ has a Gaussian process prior,
$F\sim \N(\mu,\, \Sigma)$ for some $n$-dimensional mean $\mu$ and an $n\times n$ covariance matrix $\Sigma$.
The first-order Taylor series of $G$ around an $n$-dimensional vector $m$ is $G(F) \approx G(m) + J_m (F-m)$ where
$J_m$ is the Jacobian matrix of $G(F)$ evaluated at $m$.
The corresponding approximate likelihood
from the non-linear Gaussian model ${S}^\T y \sim \N\{{S}^\T G(F),\, \sigma^2 I_{n-p} \}$
follows as
$\hat{\ddens}_m({S}^\T y\mid F) = \N\{{S}^\T y \mid {S}^\T G(m) + {S}^\T J_m (F-m),\, \sigma^2 I_{n-p} \}$, which yields the approximate posterior
\[
	\hat{\dens}_m(F \mid {S}^\T y) = \N \left(
		\Sigma^*_m
		\left[
			\frac{1}{\sigma^2} J_m^\T S S^\T \left\{
				y - G(m) + J_m m
			\right\}
			+ \Sigma^{-1} \mu
		\right]
		,\,
		\Sigma^*_m
	\right),
\]
where $\Sigma^*_m = (
			J_m^\T S S^\T J_m / \sigma^2 + \Sigma^{-1}
		)^{-1}$.
The posterior mean suggests the iterative update
\[
	m_{t+1} = (1-\rho_t)\, m_t + \rho_t\, \Sigma^*_{m_t}
		\left[
			\frac{1}{\sigma^2} J_{m_t}^\T S S^\T \left\{
				y - G(m_t) + J_{m_t} m_t
			\right\}
			+ \Sigma^{-1} \mu
		\right],
\]
where $t$ is the iteration number and $\rho_t$ the learning rate.
This update produces a dampened Gauss-Newton algorithm.
Since the mean of a Gaussian equals its mode, $m_\infty$ targets the mode of the exact posterior as $t\to\infty$.
Therefore, $\hat{\dens}_{m_\infty}(F\mid {S}^\T y)$ provides a Laplace approximation ${\hat{\law}(F \mid {S}^\T y)}$.

\begin{figure}[bp]
	\centering
	\figurebox{}{\textwidth}{}[simulation_correlated_subset_exact.eps]
	\caption{Median (dot) and interquartile ranges (x) of the absolute differences between posterior inclusion probability (PIP) and their approximation from the simulation in \Sec{}\ref{sec:IRGAsim}.
	Integrated rotated Gaussian approximation is in black,
	expectation propagation in blue,
	and variational Bayes in red.
	}
	\label{fig:cor_subset_exact}

	\figurebox{}{\textwidth}{}[simulation_correlated_subset_exact_time.eps]
	\caption{Median (dot) and interquartile ranges (x) of the computation times for the results in Fig.~\ref{fig:cor_subset_exact}.
	Integrated rotated Gaussian approximation is in black,
	expectation propagation in blue,
	and variational Bayes in red.
	}
	\label{fig:cor_subset_exact_time}
\end{figure}

\begin{figure}[bp]
	\centering
	\figurebox{}{\textwidth}{}[simulation_correlated_subset_gibbs.eps]
	\caption{Median (dot) and interquartile ranges (x) of the absolute differences between the posterior inclusion probability (PIP) approximation and their Gibbs sampler estimate from the simulation in \Sec{}\ref{sec:IRGAsim}.
	Integrated rotated Gaussian approximation is in black,
	expectation propagation in blue,
	and variational Bayes in red.
	}
	\label{fig:cor_subset_gibbs}

	\figurebox{}{\textwidth}{}[simulation_correlated_subset_gibbs_time.eps]
	\caption{Median (dot) and interquartile ranges (x) of the computation times for the results in Fig.~\ref{fig:cor_subset_gibbs}.
	Integrated rotated Gaussian approximation is in black,
	expectation propagation in blue,
	and variational Bayes in red.
	}
	\label{fig:cor_subset_gibbs_time}
\end{figure}

\section{Additional simulation studies}
\label{sec:add_simulations}

\subsection{Variable selection on a correlated subset}
\label{sec:IRGAsim}

Consider the set-up from \Sec{}\ref{sec:covariates} with the same spike-and-slab prior on the elements of $\beta$ as on the elements of $\alpha$, $n=100$, $p=2$, $\psi = 1$, $\lambda = p/(p+q)$, and $\sigma^2 = 1/2$.
Generate the elements in $X$ and $Z$ independently from $\N(0, 1)$, then reassign the second column of $X$, denoted by $X_{\ast 2}$, to equal $0.01X_{\ast 2} + 0.99 X_{\ast 1}$ to induce correlation,
and lastly standardize the columns of $X$ and $Z$ to have zero mean and unit standard deviation.
Generate $y$ according to \eqref{eq:lm}
with $\alpha = (0,\dots,0)^\T$ and $\beta = (1,2)^\T$.
Then, compute the posterior inclusion probabilities for $\beta$ using Algorithm~\ref{alg:irga} with vector approximate message passing in Step~2 as described in \Sec{}\ref{sec:covariates},
and using expectation propagation and variational Bayes as in \Sec{}\ref{sec:diabetes} but with $\sigma^2$ known.
Do this for $q=1,2,\dots,15$ with exact computation of the posterior inclusion probabilities as reference.
For large $q$, exact computation takes too long. Therefore, use a Gibbs sampler with 10,000 burnin and 90,000 recorded iterations to compute reference posterior inclusion probabilities for $q=15,30,\dots,480,960$.
Repeat the above 20 times for each $q$.

Figs.~\ref{fig:cor_subset_exact} through \ref{fig:cor_subset_gibbs_time} contain the results and computation times.
Integrated rotated Gaussian approximation has the lowest approximation error, although the difference with expectation propagation is less pronounced in Fig.~\ref{fig:cor_subset_gibbs} as approximation error from the method and Monte Carlo error from the Gibbs sampler are mixed.
Comparing $q=15$ in Fig.~\ref{fig:cor_subset_exact} and Fig.~\ref{fig:cor_subset_gibbs}
shows that the Monte Carlo error is of noticeable size compared to the approximation error of our method and expectation propagation.
Integrated rotated Gaussian approximation deals with the fact that the columns of $X$ are correlated since $\beta$ is treated separately in Algorithm~\ref{alg:irga}.
Expectation propagation and variational Bayes do not make such a distinction between the elements of $\alpha$ and $\beta$.
Variational Bayes consistently has the highest approximation error and the shortest computation time.
Expectation propagation has a higher computation time than integrated rotated Gaussian approximation which is a result of how quickly vector approximate message passing converges in this set-up.

\begin{figure}[tbp]
	\centering
	\figurebox{}{\textwidth}{}[simulation_random_design.eps]
	\caption{Median (dot) and interquartile ranges (x) of the absolute differences between the posterior inclusion probability (PIP) approximation and their Gibbs sampler estimate from the simulation in \Sec{}\ref{sec:random_mat}.
	Integrated rotated Gaussian approximation is in black,
	expectation propagation in blue,
	and variational Bayes in red.
	}
	\label{fig:random_mat}

	\figurebox{}{\textwidth}{}[simulation_random_design_time.eps]
	\caption{Median (dot) and interquartile ranges (x) of the computation times for the results in Fig.~\ref{fig:random_mat}.
	Integrated rotated Gaussian approximation is in black,
	expectation propagation in blue,
	and variational Bayes in red.
	}
	\label{fig:random_mat_time}
\end{figure}

\subsection{Variable selection with a random design matrix}
\label{sec:random_mat}

Consider the set-up from \Sec{}\ref{sec:bvs}
with $n=100$, $\lambda = 40/r$, $\psi = 1$, and $\sigma^2 = 1/2$.
Generate
$\theta$ by
randomly selecting 40 elements in $\theta$ to be non-zero and drawing them from $\N(0, 1)$. The elements of $A$ are drawn independently from $\N(0, 1)$ after which the columns of $A$ are standardized to have zero mean and unit standard deviation.
Sample $y$ according to $y\sim\N(A\theta,\, \sigma^2 I_n)$.
We repeat the random generation of $\theta$, $A$, and $y$
20 times for each $r=60,120,240,480,960$.
Estimate the posterior inclusion probabilities
using the same methods as in \Sec{}\ref{sec:diabetes}
but with $\sigma^2=1/2$ known and without considering the debiased lasso.
Algorithm~\ref{alg:irga} is used with $p = \lfloor \log(r) \rfloor$ as in \Sec{}\ref{sec:diabetes}.

The results and computation times are in Figs.~\ref{fig:random_mat}
and \ref{fig:random_mat_time}, respectively.
There is no clear separation between the methods in terms of their approximation errors in Fig.~\ref{fig:random_mat}. This might be a result of the smoothening effect of the Monte Carlo error which adds to the reported error as in Fig.~\ref{fig:cor_subset_gibbs}.
Our method seems to yield slightly more accurate posterior inclusion probabilities for $r=60,120,240$, when Monte Carlo error is also lower because $r$ is smaller, albeit at a higher computational cost.
The higher computational cost results from our method having to repeat Algorithm~\ref{alg:irga} $\lceil r/p \rceil$ times to obtain all $r$ posterior inclusion probabilities, while we use only $8\ll \lceil r/p \rceil$ CPU cores for parallelization here.
Expectation propagation and variational Bayes target all posterior inclusion probabilities at once.

\subsection{Variable selection with gene expressions}
\label{sec:leukemia}

In \Sec{}\ref{sec:random_mat}, $A$ was simulated. Let us instead set $A$ equal to 3,571 expression levels from the leukemia data from \citet{Golub1999} available in the supplementary data of \citet{Friedman2010}. Then, $n=72$ and $r=3{,}571$.
More importantly, the predictors are now highly dependent in a complex, non-linear, and non-Gaussian way: For instance, the maximum correlation between columns of $A$ equals $0.988$. 
The rest of the simulation, which we repeat 10 times, is the same as in \Sec{}\ref{sec:random_mat}.
The results are in
Table~\ref{tab:leukemia}.

Expectation propagation and our method achieve similar performance, with similar error sizes. On the other hand, expectation propagation is over twice as fast. Variational Bayes takes an order of magnitude less computation time than expectation propagation but yields worse approximations.
As in \Sec{}\ref{sec:random_mat}, the longer computation time of our method stems from having to repeat Algorithm~\ref{alg:irga} to obtain all posterior inclusion probabilities.

\begin{table}[tbp]
\caption{Summary statistics of the absolute difference between the Gibbs sampler estimates and the approximations of the posterior log odds of inclusion for the simulation study in \Sec{}\ref{sec:leukemia}
with median computation times.
IRGA stands for integrated rotated Gaussian approximation.
\label{tab:leukemia}}
\centering
\begin{tabular}{r|cccccc|c}
Method & Min & Q1 & Median & Q3 & Max & Mean & \thead{Median computation \\ time (seconds)} \\
\hline
IRGA & 0.000 & 0.231 & 0.456 & 0.713 & 50.1 & 0.540 & 54 \\
Expectation propagation & 0.000 & 0.192 & 0.404 & 0.699 & 46.9 & 0.529 & 21 \\
Variational Bayes & 0.003 & 1.50 & 1.83 & 2.25 & 56.7 & 2.44 & 1.5
\end{tabular}
\end{table}

\section{Options for splitting $\theta$ into $\alpha$ and $\beta$}
\label{sec:theta_split}

\subsection{Motivation}

Using Algorithm~\ref{alg:irga} for Bayesian variable selection as detailed in \Sec{}\ref{sec:bvs} requires repeatedly splitting the $r$-dimensional coefficient vector $\theta$ into the $q$-dimensional $\alpha$ and $p$-dimensional $\beta$.
Different splits yield different correlation structures between the columns of $X$ and $Z$. Such correlation might affect the quality of the approximation $\hat{\law}(\beta \mid y)$. Therefore, one might want to choose splits that minimize the correlation between the columns of $X$ and $Z$. Also, one can average the obtained posterior inclusion probabilities over multiple splits to reduce dependence on any one splitting of $\theta$.

This section considers multiple options for splitting $\theta$. The resulting approximation accuracy of these options is empirically compared.
For ease of exposition, we assume that $r$ is divisible by $p$.
Then, $r/p$ splits are required to compute $\hat{\pr}(\theta_j\ne 0\mid y)\ (j = 1,\dots,r)$ by repeating Algorithm~\ref{alg:irga}.
The methods are readily modified for when $r$ is not divisible by $p$ by using $\lceil r/p \rceil$ splits where in the $\lceil r/p \rceil p - r$ last splits $\beta$ is $(p-1)$-dimensional instead of $p$-dimensional.

\subsection{Methods of splitting $\theta$}

One method for splitting $\theta$ is sequential. The first split is $\alpha = (\theta_{p+1},\dots,\theta_r)^\T$ and $\beta = (\theta_{1},\dots,\theta_p)^\T$. The $k$th split is $\alpha = (\theta_{1},\dots,\theta_{(k-1)p},\theta_{kp+1},\dots,\theta_r)^\T$ and $\beta = (\theta_{(k-1)p+1},\dots,\theta_{kp})^\T$ for $k = 2,\dots,r/p-1$. The final split is $\alpha = (\theta_1,\dots,\theta_{r-p})^\T$ and $\beta = (\theta_{r-p+1},\dots,\theta_r)^\T$.

Splitting can also be done randomly. Sequential splitting depends on the ordering of the columns in the design matrix $A$.
Instead, we can randomly permute the elements of $\theta$ and the respective columns of $A$ before splitting sequentially.
This breaks the dependence on the ordering but introduces dependence on the permutation. To reduce dependence on a single permutation, one can use multiple random permutations, and then take the average of the multiple $\hat{\pr}(\theta_j\ne 0\mid y)$ obtained.
Here, we use 10 random permutations.

Sequential and random splitting do not minimize the correlation between the columns of $X$ and $Z$.
We present two options that aim to minimize this correlation.
The first option, which we call Belsley splitting, is based on \citet[\Sec{}3.2]{Belsley1980} and only applies if $r\leq n$.
Let $A = UDV^\T$ denote a singular-value decomposition with an $n\times r$ matrix $U$, an $r\times r$ diagonal matrix $D$ of singular values in decreasing order, and an $r\times r$ matrix $V$.
Then, we compute $\phi_{kj} = V_{kj}^2/D_{jj}^2$ and $\pi_{jk} = \phi_{kj} / \sum_{j=1}^p \phi_{kj}$ for $j,k = 1,\dots,r$ as in \citet[Equation~3.11]{Belsley1980}.
A larger $\pi_{jk}$ indicates collinearity between the $j$th and the $k$th column of $A$ if $j\ne k$ and each row of the matrix $\pi$ corresponds with a different potential near linear dependency between the columns of $A$ in decreasing order of severity as discussed in \citet[\Sec{}3.2]{Belsley1980}.
To group collinear columns of $A$ together, the $r/p$ splits of $\theta$ are as follows. In the $j$th split, $\beta$ consists of the $p$ elements $\theta_k$ for which $\pi_{jk}$ is the largest, and $\alpha$ consists of the other elements in $\theta$.

The second option, which we call spectral splitting, is based on spectral clustering \citep{vonLuxburg2007} and also applies if $r>n$.
Consider a weighted graph with the columns of $A$ as nodes and the absolute value of the correlation between two columns as the edge weight.
Then, the $r\times r$ similarity matrix $W$ has $W_{jk}$ equal to the absolute value of the correlation between the $j$th and the $k$th column of $A$.
Here, $W_{jk}=0$ and $W_{jk}=1$ correspond with least and most similarity, respectively, between the $j$th and the $k$th column of $A$.
The $r/p$ splits of $\theta$ follow from spectral clustering using $W$.
The Laplacian matrix of the graph is $L = D - W$ where $D$ is a diagonal matrix with $D_{kk} = \sum_{j=1}^r W_{jk}\ (k = 1,\dots,r)$.
Let the $r\times (r/p)$ matrix $U$ consist of the first $r/p$ eigenvalues of $L$.
Then, the rows of $U$ constitute $r$ points in $\mathbb{R}^{r/p}$.
We cluster the points into $r/p$ clusters using $k$-means clustering.
These clusters are not necessarily of equal size and $\beta$ cannot contain too many elements from $\theta$ as then evaluation of \eqref{eq:irga} is computationally expensive.
Therefore, we reassign points in clusters of size greater than $r/p$ to other nearby clusters such that no cluster contains more than $r/p$ points.
Each cluster corresponds with columns of $A$ and thus elements of $\theta$.
Each split of $\theta$ follows by having $\beta$ consist of the elements of $\theta$ from one cluster while the other elements in $\theta$ constitute $\alpha$.

\subsection{Empirical comparison}

This section investigates how the various options for splitting $\theta$ affect the approximation accuracy. Firstly, consider the set-up from \Sec{}\ref{sec:diabetes}. We run Algorithm~\ref{alg:irga} with vector approximate message passing, and with sequential, random, Belsley and spectral splitting.
Table~\ref{tab:diabetes_theta_split} contains the results. It shows that the approximation accuracy does not vary notably with the various methods for splitting $\theta$.

\begin{table}[bp]
\caption{Summary statistics of the absolute difference between the Gibbs sampler estimates and the approximations of the posterior log odds of inclusion for the diabetes data.
The approximations come from Algorithm~\ref{alg:irga} with different methods of splitting $\theta$.
\label{tab:diabetes_theta_split}}
\centering
\begin{tabular}{r|cccccc}
Method for splitting $\theta$ & Min & Q1 & Median & Q3 & Max & Mean \\
\hline
Sequential & 0.003 & 0.036 & 0.076 & 0.133 & 10.7 & 0.599 \\
Random & 0.005 & 0.027 & 0.061 & 0.113 & 8.66 & 0.475 \\
Belsley & 0.005 & 0.034 & 0.060 & 0.140 & 12.5 & 0.588 \\
Spectral & 0.000 & 0.032 & 0.071 & 0.139 & 9.93 & 0.601 \\
\end{tabular}
\end{table}

\begin{table}[tbp]
\caption{Summary statistics of the absolute difference between the Gibbs sampler estimates and the approximations of the posterior inclusion probability for data simulated as in \Sec{}\ref{sec:random_mat} with $r=60$.
The approximations come from Algorithm~\ref{alg:irga} with different methods of splitting $\theta$.
\label{tab:random_theta_split}}
\centering
\begin{tabular}{r|cccccc}
Method for splitting $\theta$ & Min & Q1 & Median & Q3 & Max & Mean \\
\hline
Sequential & 0.000 & 0.000 & 0.001 & 0.007 & 0.368 & 0.007 \\
Random & 0.000 & 0.000 & 0.000 & 0.006 & 0.123 & 0.005 \\
Belsley & 0.000 & 0.000 & 0.001 & 0.007 & 0.134 & 0.006 \\
Spectral & 0.000 & 0.000 & 0.001 & 0.006 & 0.130 & 0.006 \\
\end{tabular}
\end{table}

\begin{table}[tbp]
\caption{Summary statistics of the absolute difference between the Gibbs sampler estimates and the approximations of the posterior log odds of inclusion for data simulated as in \Sec{}\ref{sec:leukemia}.
The approximations come from Algorithm~\ref{alg:irga} with different methods of splitting $\theta$.
\label{tab:leukemia_theta_split}}
\centering
\begin{tabular}{r|cccccc}
Method for splitting $\theta$ & Min & Q1 & Median & Q3 & Max & Mean \\
\hline
Sequential & 0.000 & 0.253 & 0.500 & 0.834 & 52.9 & 0.769 \\
Random & 0.000 & 0.280 & 0.550 & 0.954 & 50.7 & 0.838 \\
Spectral & 0.000 & 0.247 & 0.494 & 0.828 & 51.8 & 0.760 \\
\end{tabular}
\end{table}

Secondly, repeat this comparison but now with the set-up from \Sec{}\ref{sec:random_mat} with $r=60$.
The results are in Table~\ref{tab:random_theta_split}.
Again, the approximation accuracy does not vary notably with the various methods for splitting $\theta$.

Lastly, we run the comparison on the gene expression data from \Sec{}\ref{sec:leukemia}. Here, we cannot use Belsley splitting since $r>n$ in these data.
Once more, the results in Table~\ref{tab:leukemia_theta_split} do not show notable variation in approximation accuracy across the various methods for splitting $\theta$.

\subsection{Choice of split size $p$}
\label{sec:split_size}

So far, we have used $p = \lfloor \log(r) \rfloor$ as suggested by $p=O(\log r)$ from \Sec{}\ref{sec:bvs} as a trade-off between approximation accuracy and speed.
Here, we investigate how the approximation accuracy can vary with $p$.
We run Algorithm~1 with sequential splitting and $p=1,2,4,8,16$ on the gene expression data from \Sec{}\ref{sec:leukemia}.
Table~\ref{tab:leukemia_split_size} contains the results.
Approximation accuracy is better at $p=16$. However, $p$ larger than $O(\log r)$ can increase computational cost exponentially since the cost of computing \eqref{eq:irga} is exponential in $p$ for Bayesian variable selection.

\begin{table}[tbp]
\caption{Summary statistics of the absolute difference between the Gibbs sampler estimates and the approximations of the posterior log odds of inclusion for data simulated as in \Sec{}\ref{sec:leukemia}.
The approximations come from Algorithm~\ref{alg:irga} with sequential splitting using different split sizes $p$.
\label{tab:leukemia_split_size}}
\centering
\begin{tabular}{r|cccccc}
$p$ & Min & Q1 & Median & Q3 & Max & Mean \\
\hline
1 & 0.000 & 0.244 & 0.472 & 0.730 & 55.2 & 0.608 \\
2 & 0.000 & 0.242 & 0.475 & 0.735 & 55.5 & 0.632 \\
4 & 0.000 & 0.239 & 0.466 & 0.733 & 55.1 & 0.632 \\
8 & 0.000 & 0.229 & 0.450 & 0.711 & 54.5 & 0.588 \\
16& 0.000 & 0.212 & 0.422 & 0.673 & 16.6 & 0.522 \\
\end{tabular}
\end{table}

\section{Additional comparisons for \Sec{}\ref{sec:snp}}
\label{sec:snp_supp}

This section provides additional results for the SNP application from \Sec{}\ref{sec:snp}.
In addition to integrated rotated Gaussian approximation and ignoring the SNPs, we consider the following methods for inference on $\beta$.
1) Expectation propagation from \citet{HernandezLobato2014} is readily extended to allow for different prior inclusion probabilities and slab variances per coefficient. As such, we can use it in the current set-up where the spike-and-slab prior on $\beta$ is different from the spike-and-slab prior on $\alpha$.
2) We run a Gibbs sampler with 10,000 burnin and 90,000
recorded iterations.
3) The model in \eqref{eq:lm} can be used as a mixed effects model with fixed effects $\beta$ and random effects $\eta$. Here, we fit a mixed effects model \citep{Bates2015} to exemplify this interpretation of \eqref{eq:lm} and to provide a comparison with a frequentist method. The clusters for the random effects are 17 groups of individuals that are genetically distinct according to the 2,000 SNPs with the highest sure independence screening score \citep{Fan2008}. The mixed effects model provides estimates of $\beta$ but no posterior inclusion probabilities since it is not a Bayesian method.

The resulting posterior inclusion probabilities and estimates for $\beta$ with computation times are in Tables~\ref{tab:snp_supp} and \ref{tab:snp_mean}, respectively.
The results from the Gibbs sampler and expectation propagation are consistent and different from our method.
Either our method's approximations are inaccurate, those from the Gibbs sampler and expectation propagation are, or they all are.
The Gibbs sampler and expectation propagation being consistent suggests that our method failed to give accurate approximations for this posterior. Though, it is also possible that both the Gibbs sampler and the expectation propagation struggle with the high-dimensional posterior in such a way that results in similar but inaccurate approximations.
Our method requires substantially less computation time than expectation propagation and the Gibbs sampler.

The mixed effects model yields estimates that are not aligned with the posterior means from any of the Bayesian options. This is unsurprising since the mixed effects model is adjusting for the SNPs in a fundamentally different manner.

\begin{table}[tbp]
\caption{Posterior inclusion probabilities for the demographic factors from the application in \Sec{}\ref{sec:snp}.
EP and IRGA stand for integrated rotated Gaussian approximation and expectation propagation, respectively.
\label{tab:snp_supp}}
\centering
\begin{tabular}{r|c|cccc}
& & \multicolumn{4}{c}{Population} \\
Method & Gender & Utahn of European ancestry & Finnish & Tuscan & Yoruba \\
\hline
IRGA & 0.83 & 0.96 & 0.96 & 0.92 & 0.00 \\
EP & 0.18 & 0.05 & 0.05 & 0.05 & 1.00 \\
Gibbs sampler & 0.19 & 0.05 & 0.05 & 0.06 & 1.00 \\
\end{tabular}
\end{table}

\begin{table}[tbp]
\caption{Posterior mean or estimates of $\beta$ corresponding with the demographic factors from the application in \Sec{}\ref{sec:snp}.
EP and IRGA stand for integrated rotated Gaussian approximation and expectation propagation, respectively.
\label{tab:snp_mean}}
\centering
\begin{tabular}{r|c|cccc|c}
& & \multicolumn{4}{c}{Population} \\
Method & Gender & \thead{Utahn of\\ European ancestry} & Finnish & Tuscan & Yoruba & \thead{Computation \\ time} \\
\hline
IRGA & -0.012 & -0.001 &  0.001 & 0.004 & 0.189 & 15 seconds \\
Ignoring the SNPs & -0.083 & 0.003 & 0.000 & 0.015 & -0.050 & 82 millisecs. \\
EP & -0.014 & -0.001 & 0.002 & 0.002 & 0.336 & 21 minutes \\
Gibbs sampler & -0.015 & -0.001 & 0.002 & 0.002 & 0.323 & 5.2 days \\
Mixed effects model & -0.124      &  0.010     &   0.012     &   0.031    &   -0.049 & 1.6 seconds \\
\end{tabular}
\end{table}

\small{
\bibliographystyle{biometrika}
\bibliography{biomet.bib}
}